\documentclass[aps,physrev,twocolumn,groupedaddress]{revtex4-2}
\usepackage{amssymb,amsmath}
\usepackage{xcolor}
\usepackage{lineno}
\usepackage{subcaption}
\usepackage{graphicx}

\begin{document}

%\preprint{}

\title{Optical Echo State Network Reservoir Computing}


\author{Ishwar S. Kaushik}
\email[]{Contact author: ishwarskaushik@arizona.edu}
\author{Peter J. Ehlers}
\author{Daniel Soh}
\email[]{Contact author: danielsoh@arizona.edu}
%\thanks{}
\affiliation{Wyant College of Optical Sciences, University of Arizona, Tucson, Arizona, USA}


\date{\today}

\begin{abstract}
We propose an innovative design for an optical Echo State Network (ESN), an advanced type of reservoir computer known for its universal computational capabilities. Our design enables an optical implementation of arbitrary ESNs, featuring flexibility in optical matrix multiplication and nonlinear activation. Leveraging the nonlinear characteristics of stimulated Brillouin scattering (SBS), the architecture efficiently realizes measurement-free nonlinear activation. The approach significantly reduces computational overhead and energy consumption compared to traditional software-based methods. Comprehensive simulations validate the system's memory capacity, nonlinear processing strength, and polynomial algebra capabilities, showcasing performance comparable to software ESNs across key benchmark tasks. Our design establishes a feasible, scalable, and universally applicable framework for optical reservoir computing, suitable for diverse machine learning applications.
\end{abstract}

%\keywords{}

\maketitle

\section{Introduction}
\label{sec:introduction}

 Recurrent neural networks (RNNs) have inherent memory capabilities, and are trained on sequential or time series data to create a machine learning model that attempts to make accurate time series predictions \cite{hochreiter1997long}. However, their reliance on multiple layers of tunable weights leads to high computational and training costs. A potential way to overcome this cost is with a recent paradigm of RNNs with fixed internal weights, known as reservoir computers (RCs). Reservoir computing has been of recent interest, with applications to time-series prediction tasks, classification tasks, and much more \cite{yan2024emerging,gauthier2021next,pathak2018model,vlachas2020backpropagation,zimmermann2018observing,canaday2018rapid,wang2024ultrafast,ehlers2025stochastic,ehlers2025improving,zhu2024practical,zhu2024minimalistic}. Unlike RNNs, which require large amounts of time and resources, RCs offer a much less computationally expensive method for machine learning.

An RC is a nonlinear dynamical system driven by time-dependent inputs and a single trainable output layer. Unlike RNNs or deep learning architectures, the inner layers of the computing reservoir are fixed and all of the training takes place in the final output layer. This output is usually a simple linear combination of the values of the state-vector of the reservoir, which allows for the weights to be determined by simple linear regression. This property makes reservoir computing an attractive choice for hardware designs; manipulating the weights in multiple layers for a typical RNN in hardware is exceedingly difficult in comparison to this simpler approach. Moreover, by adopting a lean, hardware-centric design in RCs, the overall hardware overhead is significantly reduced compared to software-based systems that depend on traditional computing layers—such as CPUs, long-term and short-term memories, and their interconnects. Operating on specialized hardware further increases speed by eliminating unnecessary data processing steps, thus improving overall efficiency. 

Optical platforms for computation and machine learning can, in principle, achieve more efficient computation in terms of energy and time \cite{shastri2021photonics,wang2022optical}. One important challenge in optical approaches is how to realize the nonlinear activation. For this, various nonlinear optical processes have been studied including the nonlinear behavior of SBS \cite{phang2023photonic}. It is important to note that, while the implementations of optical reservoir computers have been reported \cite{duport2012all}, the realization of an optical RC with universal computation capability (i.e., universality) has not yet been demonstrated to our best knowledge.

Universality refers to a computing machine’s ability to perform any computational task with any desired accuracy. In a supervised learning context, it specifically denotes the capacity to approximate any input-output relationship with arbitrary precision. Consequently, having a mathematically rigorous proof of universality is very desirable, as it ensures that the computing scheme can robustly handle a wide range of tasks. Arbitrary RCs are not guaranteed to obey the conditions necessary to be capable of universal computations. We present, for the first time to our best knowledge, a new approach representing a novel optical implementation of Echo State Networks (ESNs), which have recently been shown to form a universal approximating class of RCs \cite{grigoryeva2018echo}. 

ESNs are a family of RCs that are formed from iteratively applying a nonlinear transformation after a fixed linear transformation applied at each input time step \cite{jaeger2001echo}. ESNs are similar to RNNs except that the linear weights given to the state of the system at each time step are fixed for the entire run. The family of ESNs as a whole is universal \cite{grigoryeva2018echo} which means that for a given target sequence and input, we can find a specific ESN configuration that can approximate the target function to any desired accuracy.

Here, we propose an innovative way to realize an optical implementation of any desired ESN configuration. The advantages provided by an optical reservoir architecture, as stated earlier, are fast computation and better energy efficiency than large software alternatives. For this reason, there has been growing interest in recent years in optical implementations of the matrix multiplication stage of machine learning architectures \cite{zhou2022photonic,xu202111,yang2013chip}. Here, we propose an in-fiber method for optical matrix-vector multiplication, which is followed by an in-fiber solution for optical nonlinear activation inspired by Slinkov et. al \cite{slinkov2024all} using SBS. This unique combination of optical matrix multiplication and nonlinear activation allows for a feasible design of the optical ESN.

The paper is organized as follows: Section \ref{sec:method} covers the basic concepts of ESNs and the theory behind the SBS linear amplifiers and activation function, Section \ref{sec:model} addresses our proposed design for the optical ESN, and Section \ref{sec:results} compiles the performance of the ESN on various tasks that evaluate its universality, and compares it to a software ESN of the same size. The tunability of the nonlinearity of SBS interaction is also analyzed in this section. Finally, Section \ref{sec:conclusion} contains the concluding remarks.

\section{Optical Reservoir Computing Concepts}
\label{sec:method}

\subsection{Echo State Network}

An ESN is a form of reservoir computing, responding to time-series inputs reflected in its state vector after each time step. The state of the reservoir at each time step is given by

\begin{equation}
    x_{k+1} = g(x_k,u_k).
    \label{eqn:rceq}
\end{equation}

Where the vector $x_k$ represents the state of the reservoir at time step $k$. The number of elements of the state vector is equal to the number of nodes $N$ of the reservoir computer. The function $g(x_k,u_k)$ is a fixed nonlinear function of the reservoir that provides the updated state of the reservoir dependent on the $k$th element of the input sequence $u_k$. The predicted output of the network is given by

\begin{equation}
    \hat{y}_k = W^T x_k + C,
    \label{eqn:yhat}
\end{equation}
and the cost function is
\begin{equation}
    S = \frac{1}{2N}\sum^{N-1}_{k=0}(y_k - \hat{y}_k)^2.
    \label{eqn:cost}
\end{equation}
The parameters $W$ and $C$ given in Equation \eqref{eqn:yhat} are chosen to minimize the cost function $S$ given in Equation \eqref{eqn:cost}. Here, $y_k$ denotes the target sequence the network output is fit to. So, the fitting of the output function to the target data is a linear regression problem. The reservoir dynamics are a nonlinear function of the inputs $u_k$. The readouts from the reservoir can be used to approximate the function $y_k$ with linear regression.

A reservoir computer is required to satisfy the uniform convergence property \cite{grigoryeva2018echo}. This property asserts that the reservoir state converges to a unique sequence after many time steps, which means that the initial state of the reservoir from the distant past has no effect on the current state. This property can be extended by including the continuity of $g(x_k,u_k)$ and result in the fading memory property \cite{boyd1985fading}. The fading memory property implies that the dependence of the state vector of the reservoir $x_k$ on the input $u_{k_0}$ at a time step $k_0$ must diminish as $k-k_0$ tends to infinity. This allows for a universal, repeatable, and deterministic reservoir while still maintaining some memory.

An ESN \cite{jaeger2001echo} is a type of reservoir computer with a form given by
\begin{equation}
    g(x_k,u_k) = f(Ax_k + Bu_k),
    \label{eqn:ESN}
\end{equation}
where $A$ is a random but fixed matrix and $B$ is a random, fixed vector. The function $f(z)$ is a nonlinear activation function that acts on the input $z = Ax_k + Bu_k$. For $N$ nodes of the reservoir, the matrix $A$ has dimensions $N$ x $N$ and the vector $B$ has dimensions $N$ x $1$. Uniform convergence of an ESN is guaranteed by the condition that $A^TA < a^2I_n$ where $a$ is a number determined by the nonlinear function $f(z)$ \cite{ehlers2023improving}. Hence, the fading memory property is also guaranteed for a stable ESN, which leads to the universal approximation capability of this class of reservoir computers.

\begin{figure*}[!tb]
     \centering
     \begin{subfigure}[b]{0.35\textwidth}
         \includegraphics[width=\linewidth]{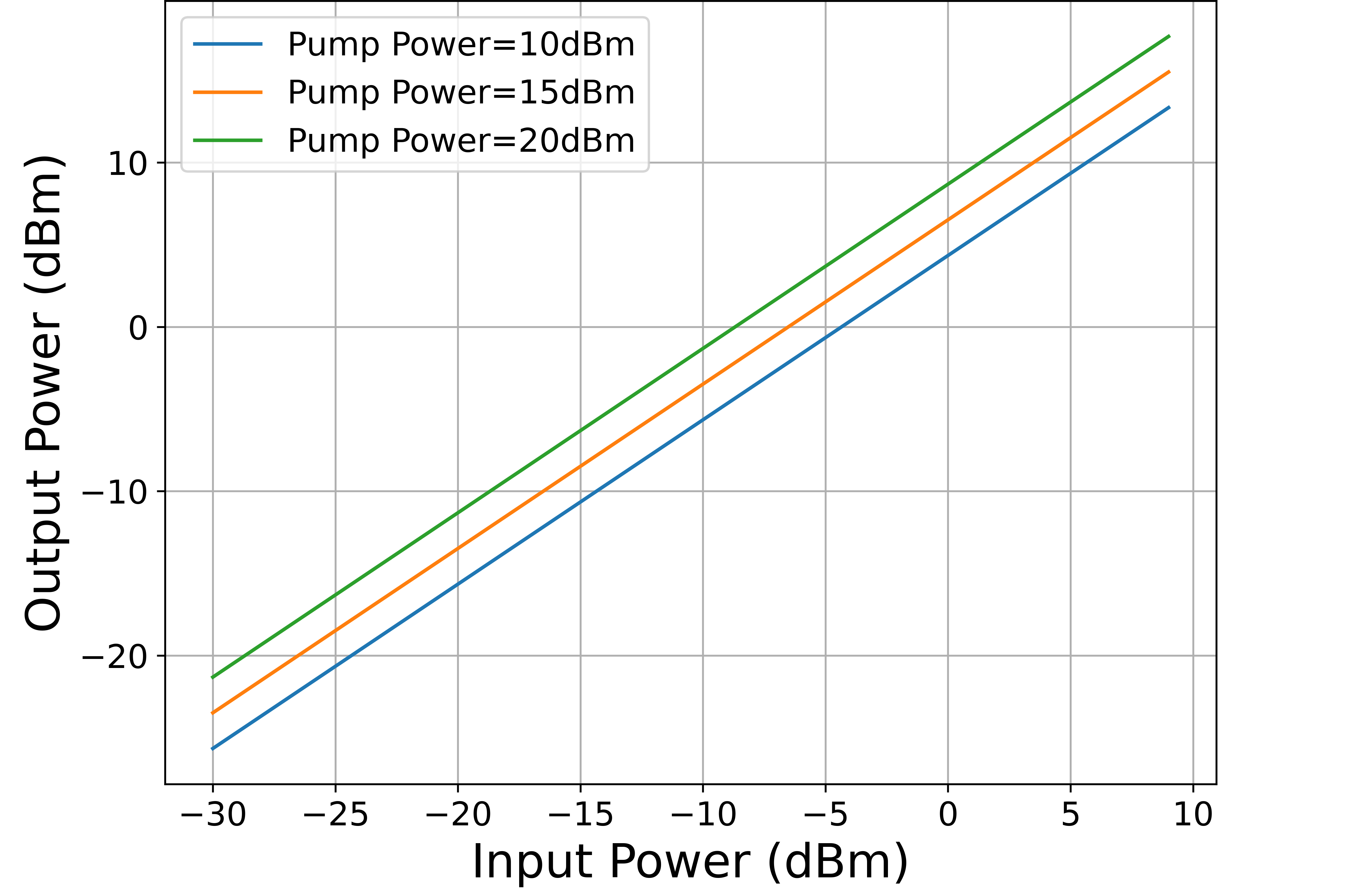}
         \caption{SBS Amplifier}
     \end{subfigure}
     \begin{subfigure}[b]{0.35\textwidth}
         \includegraphics[width=\linewidth]{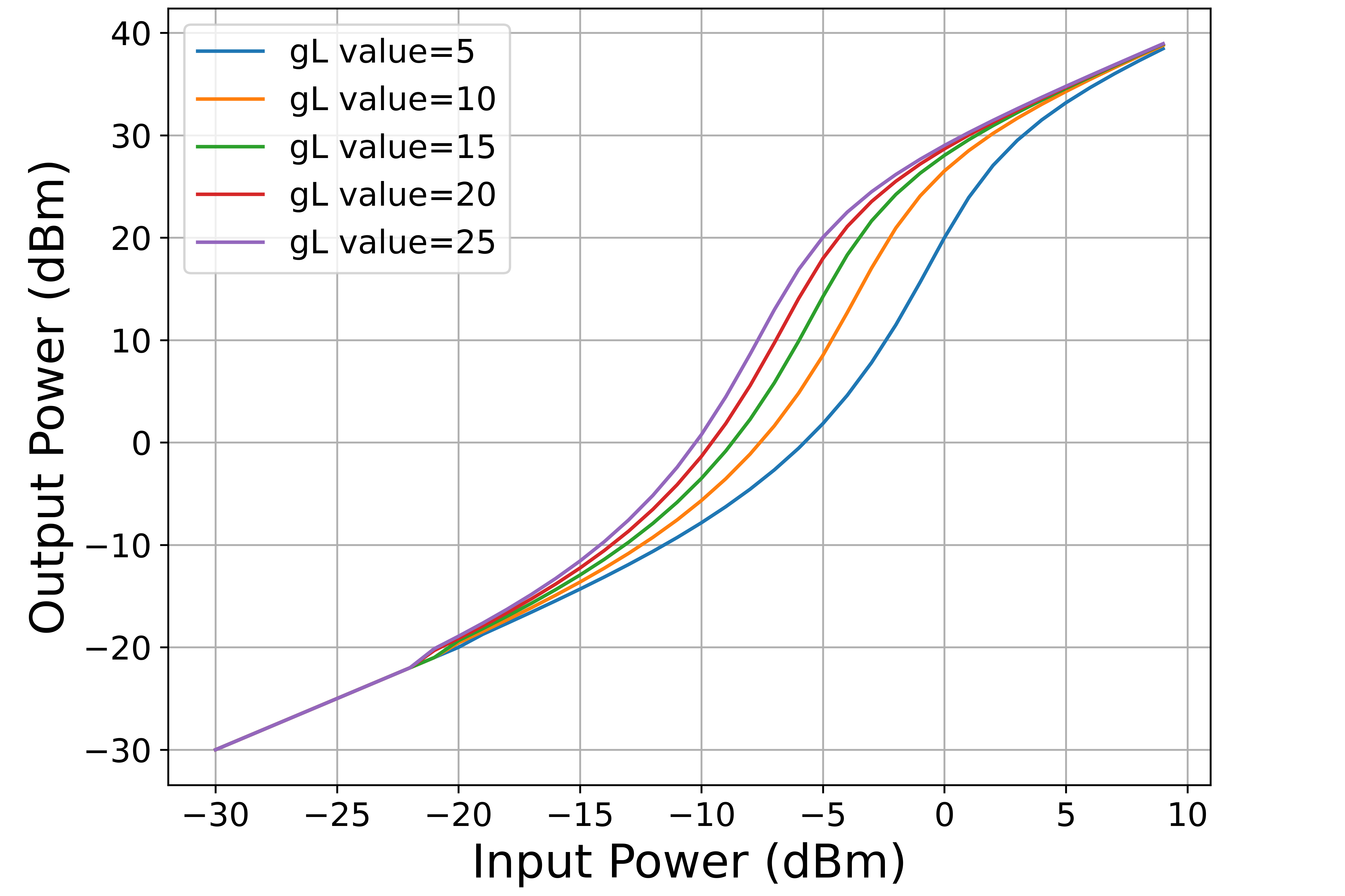}
         \caption{SBS Activation Function}
     \end{subfigure}
     \hfill
     \begin{subfigure}[b]{0.35\textwidth}
         \includegraphics[width=\linewidth]{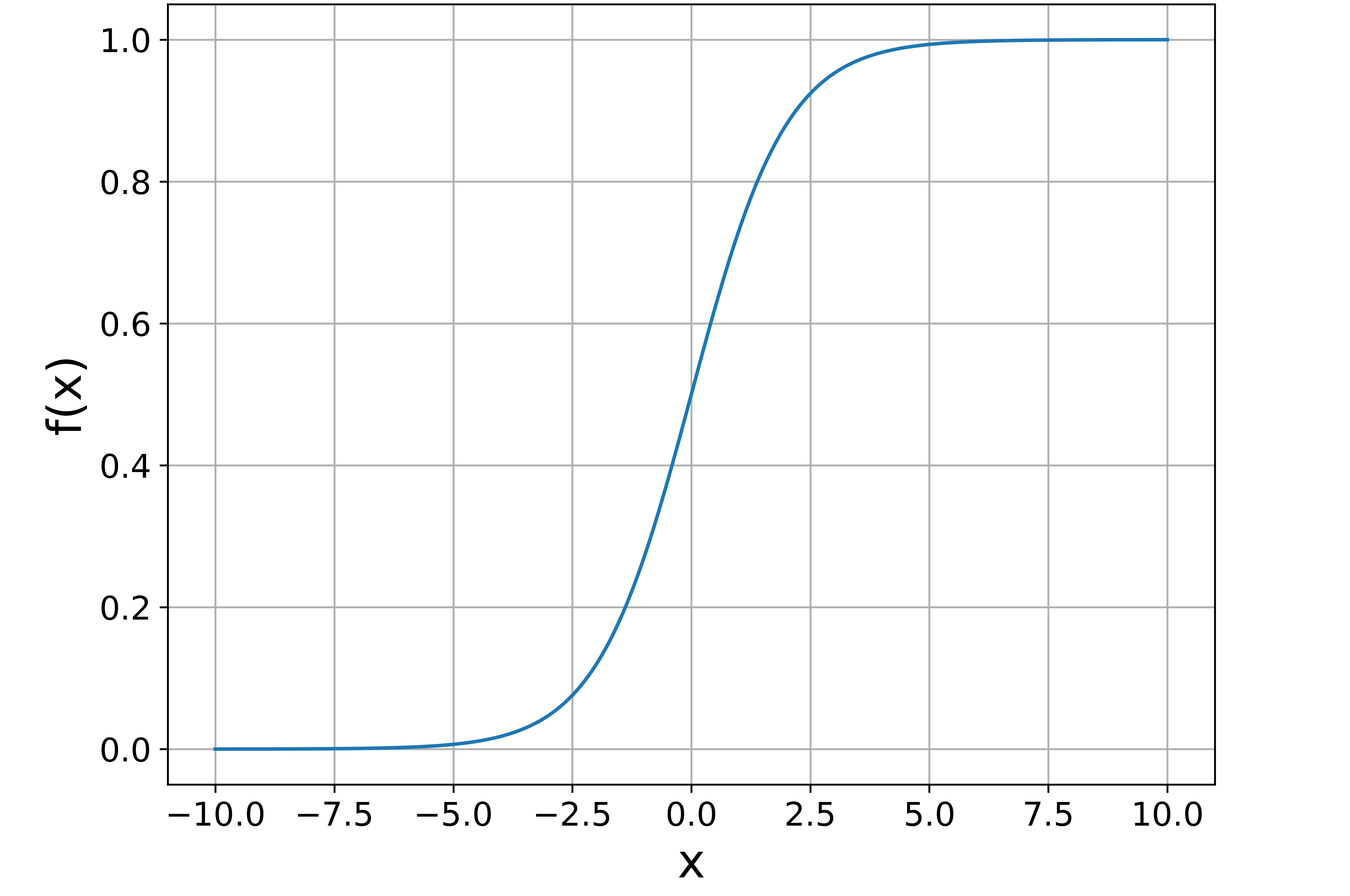}
         \caption{Sigmoid Function}
     \end{subfigure}
     \caption{(a) SBS operating in large pump (linear) regime. Input-output relationship is shown for various pump powers to demonstrate tunability. (b) SBS operating in self-pumped (nonlinear) regime. The nonlinear shape can be designed by choosing an appropriate $gL$ value. (c) Sigmoid function for comparison.}
     \label{fig:sbslinnonlin}
\end{figure*}

\subsection{Nonlinear Activation Mechanism}

SBS is a third-order nonlinear scattering process. The optical pump wave interacts with the acoustic wave of the medium, which is stimulated by the presence of the pump and probe waves. This interaction amplifies the probe wave by depleting the pump. Writing the SBS interaction in terms of coupled powers of the pump and probe optical waves \cite{boyd2008nonlinear}, we have

\begin{align}
    \frac{dP_{\mathrm{p}}}{dz} &= -gP_{\mathrm{p}}P_{\mathrm{s}}, \label{eqn:sbsp1} \\
    -\frac{dP_{\mathrm{s}}}{dz} &= +gP_{\mathrm{s}}P_{\mathrm{p}}.\label{eqn:sbsp2}
\end{align}

Equations \eqref{eqn:sbsp1} and \eqref{eqn:sbsp2} show the coupled power relations with propagation distance. The pump wave is represented by $P_{\mathrm{p}}$ and the probe wave is represented by $P_{\mathrm{s}}$. The SBS gain factor $g$ is known in terms of the Brillouin frequency, Brillouin linewidth, and other material properties of the medium like the refractive index. Solving Equations \eqref{eqn:sbsp1} and \eqref{eqn:sbsp2}, we get a transcendental equation in terms of $P_{\mathrm{out}}$ given by (see the derivation in Appendix \ref{sec:A})

\begin{equation}
    P_{\mathrm{out}}(P_{\mathrm{in}} + P_{\mathrm{p}} - P_{\mathrm{out}}) - P_{\mathrm{in}}P_{\mathrm{p}}e^{-(P_{\mathrm{p}} - P_{\mathrm{out}})gL} = 0.
    \label{eqn:pouttrans}
\end{equation}

We have defined $P_{\mathrm{s}}(L) = P_{\mathrm{out}}$. We also define $P_{\mathrm{s}}(0) = P_{\mathrm{in}}$. Here, $L$ is the length of SBS interaction. Equation \eqref{eqn:pouttrans} does not have an analytical solution but is solved approximately using numerical methods. Within the large pump approximation, which means that the pump power is much larger than the probe, the solution is linear and controlled by the pump power. When the pump power is made dependent on the input power the solution is nonlinear as shown by Slinkov et. al \cite{slinkov2024all}. By amplifying and splitting the input with an appropriate ratio, we are able to optimize the nonlinear activation. We define this parameter as the ``pump splitting ratio" $m$, given by
\begin{equation}
    m = \frac{P_{\mathrm{p}}}{P_{\mathrm{in}}}.
    \label{eqn:psplrat}
\end{equation}

We use these two behaviors to encode matrix multiplications and nonlinear activation in the optical ESN implementation. The behavior of the transcendental Equation \eqref{eqn:pouttrans} has been simulated numerically, and is shown in Figure \ref{fig:sbslinnonlin}.  Particularly,  the nonlinear behavior of SBS looks very similar to a sigmoid function, which is often used as an activation function for machine learning applications. The sigmoid function is plotted for comparison in Figure \ref{fig:sbslinnonlin}(c). The linear behavior is tunable by varying the pump power and the nonlinear regime shows different behavior for different values of the $gL$ product given in the exponent of Equation \eqref{eqn:pouttrans}, and also for different values of the pump splitting ratio $m$, given by Equation \eqref{eqn:psplrat}. This allows for an optimizable activation function, which we expand on later in the paper.

\section{Optical ESN Model}
\label{sec:model}

\begin{figure*}[!tb]
    \centering
    \includegraphics[width=0.7\linewidth]{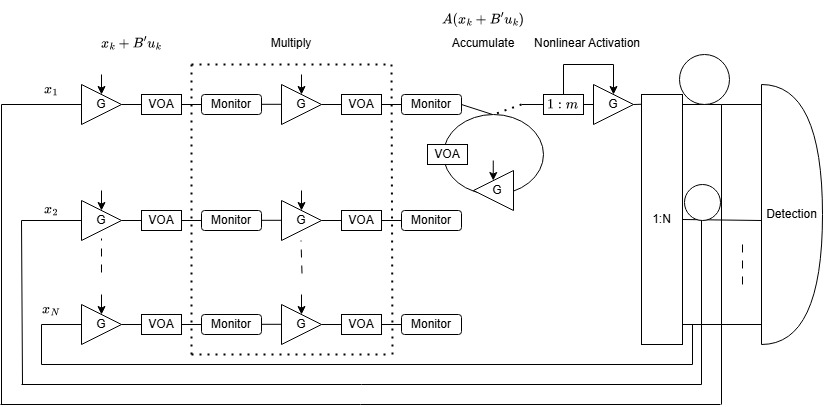}
    \caption{Optical ESN Design}
    \label{fig:optesn}
\end{figure*}

The optical ESN model is shown in Figure \ref{fig:optesn}. It is designed in an all-fiber platform to make use of highly nonlinear fiber-based SBS for the optical activation function. In order to design the ESN update Equation \ref{eqn:ESN}, the required operations are addition and multiplication, followed by nonlinear activation. The nonlinear step is achieved by the self-pumped Brillouin amplification scheme as discussed above.

We employ dBm units instead of linear units. In our algebra, all the power is converted to dBm units through
\begin{equation}
    P(\mathrm{dBm}) = 10 \log_{10} (P(\mathrm{mW})/1\mathrm{mW}).
\end{equation}
For example, if we want to perform the following algebra in our unit:
\begin{equation}
    s_\mathrm{out} = x  + u' ,
\end{equation}
where $x$ is an arbitrary dimension vector and $u'$ is a vector with the same dimension as $x$, we accomplish the above algebra in dBm units through the following physical realization:
\begin{equation}
    s_\mathrm{out} (\mathrm{mW}) = x (\mathrm{mW}) \cdot G, \quad G = 10^{0.1 u'},
\end{equation}
where the operator $\cdot$ is the element-wise multiplication and $G$ is realized through the combination of Brillouin amplifiers and the Variable Optical Attenuators (VOAs) with the gain and the loss prescribed through the vector $u$. For the case of Brillouin amplifiers, $G$ is the tunable Brillouin gain, controlled by the pump power. In the large pump approximation applied to the transcendental Equation \ref{eqn:pouttrans}, it takes the form
\begin{equation}
    G = \mathrm{exp}(gLP_{\mathrm{pump}}).
\end{equation}
$G$ can also be set to negative values with the help of a VOA. Hence, we can adjust $P_\mathrm{pump}$ value to accomplish $G = 10^{0.1 u'}$ (specifically, $P_\mathrm{pump} (u') = (1/gL) (0.1 \ln 10) u')$.

The multiplication step uses somewhat different physical mechanism, with the help of an in-line optical power monitor. As an example, if we have a specific input to output relation for this matrix multiplication given by $s_{i} = A_{ij}x_{j}$, where $x_j$ is $-1$ dBm and we wish to multiply $A_{ij} = 2$, we would set our Brillouin gain to zero gain, and set the VOA to attenuate the signal in order to achieve an output $s_i$ of $-2$ dBm. If the input $x_i$ is $+1$ dBm and we wish to multiply still the same value $A_{ij} = 2$, we would set the VOA attenuation to zero attenuation, and set the Brillouin gain to amplify the signal to make $s_i = +2$ dBm. Hence, for given $A_{ij}$ parameter, implementing the appropriate combination of the required Brillouin gain and the amount of VOA attenuation accomplishes the multiplication operation in dBm units.

The full design is shown in Figure \ref{fig:optesn}. The reservoir state vector $x_k$ (in dBm units) is implemented by the optical signal running through the optical fibers in parallel. The input $u_{k}$ is multiplied by the $B'$ vector such that $AB' = B$. Using our physical realization of the dBm unit addition scheme explained above, the vector $B'u_{k}$ is added to the state vector by appropriately tuning the gain (or loss) of the Brillouin amplifiers (or VOAs). Please note that the combination of Brillouin amplifiers and VOAs implements both the addition and the subtraction of optical powers in dB units. This is followed by the matrix multiplication step:
\begin{equation}
    s_\mathrm{out} (\mathrm{dBm}) = A s_\mathrm{in} (\mathrm{dBm}),
\end{equation}
where $s_\mathrm{in}$ and $s_\mathrm{out}$ are vectors with dimension $N$ (the dimension of the state vector $x$), and $A$ is the matrix. This operation requires the following operation for each $i$'th entry of the vector:
\begin{equation}
    s_\mathrm{out} (i) (\mathrm{dBm}) = \sum_{j=1}^N A_{ij} s_\mathrm{in} (j) (\mathrm{dBm}),
\end{equation}
which is the combination of the dBm unit addition and the dBm unit multiplication. We use the same device pairs of Brillouin amplifiers and VOAs while, first, performing the $A_{ij} s_\mathrm{in} (j) (\mathrm{dBm})$ multiplication operation in dB unit for all different $j$'s in parallel, and adding the $A_{ij} s_\mathrm{in} (j)$ in dBm units for all the $i$'s, one by one sequentially. Hence, we will require $N$ number of pairs of a Brillouin amplifier and a VOA for performing the $A_{ij} s_\mathrm{in} (j)(\mathrm{dBm})$ multiplications and one pair of a Brillouin amplifier and a VOA for adding $A_{ij} s_\mathrm{in} (j)$ in dBm units for all $j$'s sequentially.

We note that the accumulating addition steps are performed by switching the signal into a loop of fiber with the Brillouin amplifier and VOA pair, performing $\sum_j$ operation in individual time steps. This is followed by a single nonlinear activation element, to achieve each value of the output state vector. Thanks to this operation including switching into an output array of $N$ fibers with appropriate delays, the outputs from the matrix multiplication steps will be appropriately lined up into the next loop of the optical reservoir. The total number of the pairs of the Brillouin amplifier and a VOA is therefore $N+1$ as each $i$ entry requires 1 pair for dBm unit multiplication and one pair is used repeatedly for dBm unit addition.

After each row is accumulated, the resulting output is sent to the single nonlinear activation element. The pump is split off of the input with a ratio $1:m$ where $m$ is the pump splitting ratio defined by Equation \eqref{eqn:psplrat}. This implements the nonlinear activation function as shown in Figure \ref{fig:sbslinnonlin}. Any required compensatory amplification can also be achieved in the accumulation step before the nonlinear activation element. After the nonlinear activation, it is switched into the appropriate output fiber using a $1:N$ fiber switch. These outputs must be delayed to get an appropriately lined up parallel state vector. By tuning the pump splitting ratio $m$ and the $gL$ value of the self-pumped SBS function, the performance of the ESN can be optimized. This is explored in detail in Section \ref{sec:results}. 

To make this dBm scale matrix multiplication architecture more concrete, we provide a simple $2\times2$ dimensional example. If the input vector $s_{\mathrm{in}}(=x_k+B'u_k)$ is given as $\left( \begin{smallmatrix} 1 \\*[1mm] 2\end{smallmatrix} \right)$ and the desired $A$ matrix is $\left(\begin{smallmatrix}1&-2\\*[1mm]3&-1\end{smallmatrix} \right)$, the matrix multiplication $A s_\mathrm{in}$ would be performed as follows: First, for the multiplication $\left(\begin{smallmatrix} 1 & -2\end{smallmatrix} \right) \left( \begin{smallmatrix} 1 \\*[1mm] 2\end{smallmatrix} \right)$, the in-line power monitors would be used to determine the input vector powers in dBm units. The first row $\left(\begin{smallmatrix} 1 & -2\end{smallmatrix} \right)$ is multiplied by the input vector $\left( \begin{smallmatrix} 1 \\*[1mm] 2\end{smallmatrix} \right)$ by allowing the 1 dBm power in the first node to pass unchanged realizing 1 dBm $(=1 \times 1)$, and implementing $-6$ dBm  of loss to the incoming 2 dBm power in the second node realizing $-4$ dBm $(=-2 \times 2)$. The output values of $(1,-4)$ are detected by the second set of in-line power monitors. In the next step, taking place in a fiber loop accessed by a fiber switch, the addition of $1$ and $-4$ occurs through the dBm unit addition scheme that we explained above, by implementing $-4$ dB loss to the VOA on the incoming 1 dBm power. This process produces $-3$ dBm power. This first row output is allowed to pass by the fiber switch to the nonlinear activation step and switched into the first output row by the $1:N$ switch, where it encounters a delay loop. For the second row operation realizing $\left(\begin{smallmatrix} 3 & -1\end{smallmatrix} \right) \left( \begin{smallmatrix} 1 \\*[1mm] 2\end{smallmatrix} \right)$, the first node of 1 dBm power is amplified by 2 dB to realize $3$ dBm $(=3 \times 1)$ and the second node of power 2 dBm is attenuated by $-4$ dB to realize $-2$ dBm (=$(-1) \times 2)$. The output values of $(3,-2)$ are now realized in dBm scale. Then, the next step realizes the dBm unit addition of 3 dBm and $-2$ dBm. For this, a Brillouin amplifier gain of 3 dB amplifies the incoming $-2$ dBm power, which results in 1 dBm $(=3 + (-2))$ power. This output is similarly passed to the nonlinear activation step and switched to the output, where it emerges from the architecture synchronized with the first vector element computed earlier. Overall, we accomplished the matrix multiplication $\left(\begin{smallmatrix} 1 & -2 \\*[1mm] 3 & -1\end{smallmatrix} \right) \left( \begin{smallmatrix} 1 \\*[1mm] 2\end{smallmatrix} \right) = \left( \begin{smallmatrix} -3 \\*[1mm] 1 \end{smallmatrix} \right)$ in dBm units, and passed the output of the matrix multiplication through the optical nonlinear activation function.

This design allows for the choice of arbitrary matrix elements for the matrix $A$ and vector $B$, including negative elements in dBm units through the use of the appropriate attenuation and Brillouin gain. If chosen according to the convergence properties of an ESN, this design can be used to implement an arbitrary ESN, which is a universal family of reservoir computers \cite{grigoryeva2018echo}. This means that given a desired accuracy for a task (such as prediction), this method can be used to realize an optical hardware that achieves that accuracy. The proposed design also allows for high scalability, as the number of nodes is increased simply by increasing the number of parallel fibers running in the optical fiber reservoir, and the architecture uses only a single nonlinear activation element through the use of fiber switched delays. Although, scalability is naturally limited by available resources and energy.

The loop delay of each time step must also be sufficient to allow the Brillouin gain switching. As a result, this is limited by the response speed of the Brillouin amplifiers to the modulation of pump power, as well as the optical attenuation response of the VOA. Commercial-off-the-shelf electro-optic modulators (EOMs) can modulate in the tens of GHz range, allowing for quick response time of optical power variation. If we assume a 1 MHz relatively slow modulation speed for an EOM, this implies an increase in delay of 1 $\mu$s for every node. While this introduces a slow down in the speed of calculation, this tradeoff allows for a relatively simple implementation of multiply and accumulate operations in dBm scale. This can also be implemented by duplicating the power in each path and accumulating with a large number of Brillouin amplifiers, overcoming the large delay from our current approach, but this would result in a much larger energy budget due to a large number of Brillouin amplifiers, as well as a difficulty in scaling the structure. Another alternative would be to parallelize the accumulation step by accumulating multiple rows of multiplication simultaneously, greatly reducing the delay. However, this would also cause additional complexity in design and require more resources.

This also requires the SBS process to respond quickly enough to modulations of the pump. The realistic gain modulation speed of fiber Brillouin amplifiers by modulating the pump power is typically limited to tens of MHz (e.g., up to around 50–100 MHz in silica fibers). This limitation arises from the intrinsic response time of the SBS process, which is governed by the phonon lifetime (approximately 10 ns) or equivalently the Brillouin gain linewidth (50–100 MHz). For example, recent advances in optical imaging \cite{shaashoua2024brillouin} have made use of the quick response time of Brillouin amplifiers by modulating at over 1 MHz. Modulating the pump amplitude faster than the limit set by the Brillouin gain linewidth would not allow the acoustic wave (phonons) to fully respond, resulting in reduced or ineffective gain modulation.

Another important consideration is the energy consumption of the system. While an arbitrary ESN can be implemented with this architecture, the power flowing through the system can approach unfeasible amounts. In order to limit this, the scaling criteria on the amplifiers and nonlinear activation is made stringent to ensure that the maximum value of $x_k$ in the system is $31$ dBm. The maximum pump power for the nonlinear activation required in this scenario is $3.2$ W for a pump splitting ratio of $m=1000$. The pump splitting ratio is discussed further in Section \ref{sec:results}. While this results in a slight degradation in performance, we must accept this tradeoff to keep the power budget low.

Another important trade-off is between scalability and speed of operation involved. The current approach uses sequential time delays to simplify the hardware requirements. Alternatively, the architecture can be modified to fix the gain of the Brillouin amplifiers, with appropriate attenuation beforehand to always maintain convergence. This will result in a fixed reservoir implementation, but will no longer represent an ESN as the $A$ matrix will change at every time step due to a lack of correction for the input optical power. However, this comes at the cost of the guarantee of universal applicability for an ESN, therefore we have decided against this approach despite the relative hardware simplicity.

The sequential delay solution adds a different kind of hardware complexity despite reducing the required number of Brillouin amplifiers. While this is a drawback of using a fiber-based approach, we consider this fiber-based architecture to be a preliminary step towards an on-chip all optical architecture. Choudhary et al. \cite{choudhary2016advanced}, and Yu et al. \cite{yu2025chip} have shown $>50$ dB of Brillouin amplifier gain in photonic chip and $>8$ dB of gain in thin-film lithium niobate solutions respectively, showing promise that a high-gain and tunable Brillouin amplifier for the construction of a similar on-chip optical ESN may be feasible in the near future. This future photonic-circuit implementation will ensure the scalability of the machine size. 

\section{Results and Discussion}
\label{sec:results}

The optical ESN architecture was tested in simulation assuming $N$ computational nodes, which refer to $N$ elements of the $x_k$ vector. The number of nodes was varied during testing, which is expanded on below. The performance of our optical ESN was also compared with a similar ESN implemented with code in Python, using a sigmoid activation function. This comparison is intended to validate our optical ESN, not to show that one approach is superior to the other.
\begin{figure*}[!tb]
     \centering
     \begin{subfigure}[b]{0.35\textwidth}
         \includegraphics[width=\linewidth]{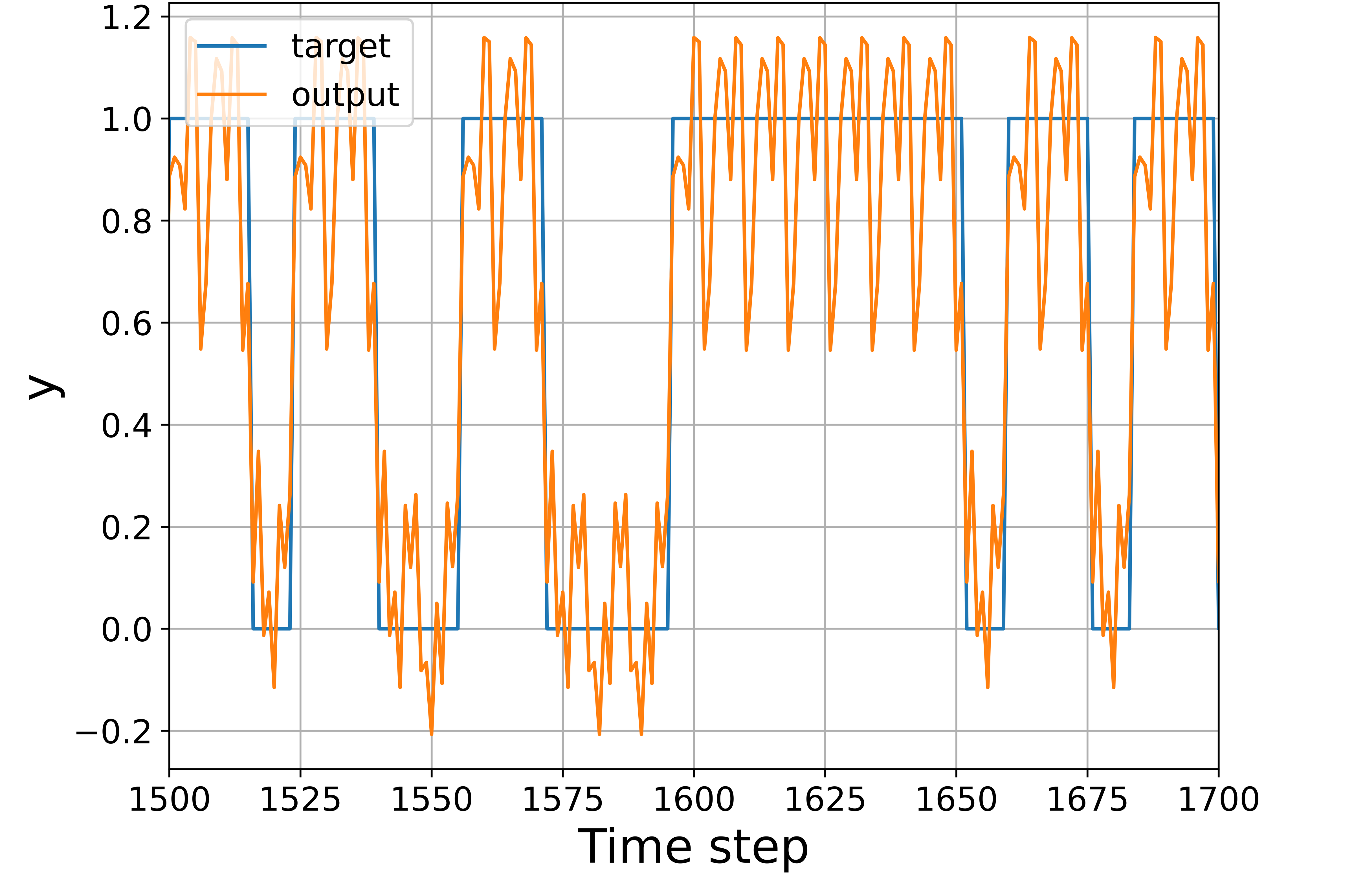}
         \caption{Software (Sigmoid) Sine Square Classification, NMSE=0.14.}
     \end{subfigure}
     \begin{subfigure}[b]{0.35\textwidth}
         \includegraphics[width=\linewidth]{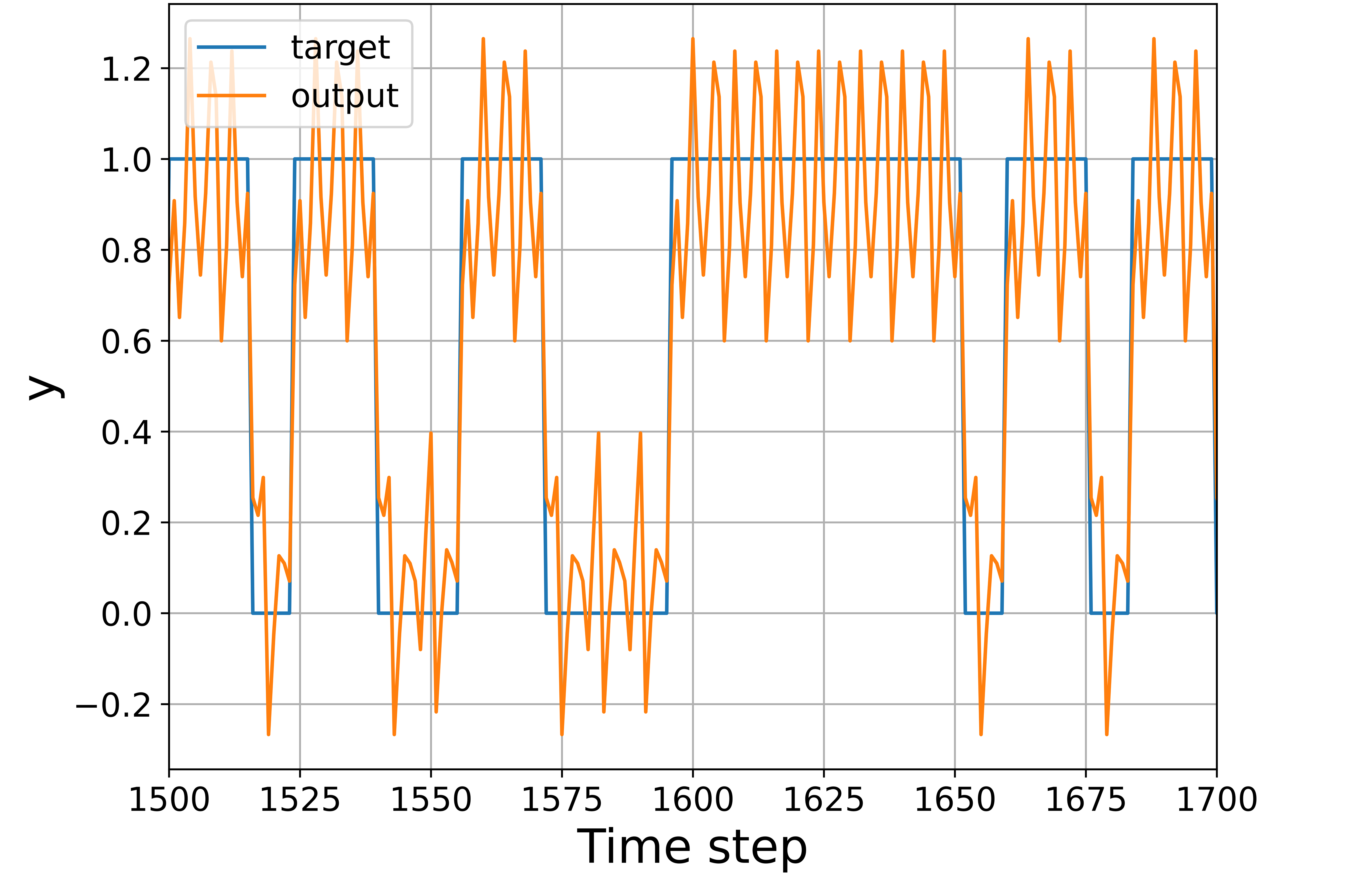}
         \caption{Optical Sine Square Classification, NMSE=0.12.}
     \end{subfigure}
     \caption{Comparison of the performance of the Sine-Square Classification test. Parameters: $N=10$, $gL=12$, $m=1500$}
     \label{fig:sinsq}
\end{figure*}
Good performance in machine learning hinges on the ability to approximate nonlinear input–output relationships and capture complex temporal dynamics. According to the \linebreak Stone–Weierstrass theorem \cite{stone1948generalized}, the polynomial algebra property is central to achieving universality. Therefore, we selected three distinct tasks to evaluate these critical capabilities. They are the sine-square classification test, prediction of the NARMA10 system, and the Mackey-Glass time series prediction. Each task evaluates mainly the reservoir computer nonlinearity, polynomial algebra capabilities, and dynamic memory retention, respectively. Additionally, we address the nonlinear properties of the SBS interaction and, thus, the optimization of the activation function. To achieve good results in the tasks mentioned above, we require strong enough nonlinearity, while maintaining the convergence conditions of the ESN. We vary the value of the $gL$ product as well as the pump splitting ratio $m$ with these constraints in mind to optimize the performance of the ESN in the selected benchmark tests. 

To evaluate the performance of the ESN, we use the normalized mean-square error (NMSE) given by
\begin{equation}
   \mathrm{NMSE} = \frac{1}{\sigma^2_y} \langle(y-\hat{y})^2\rangle = \frac{1}{N\sigma^2_y}\sum_{k=0}^{N-1}(y-\hat{y})^2.
   \label{eqn:nmse}
\end{equation}
Here, we average over $N$ time steps to represent the training interval. The estimated output $\hat{y}_k$ is given by Equation \eqref{eqn:yhat}. Here, $\sigma^2_y$ is the variance of the output $y$. So, this bounds the NMSE by the variance of $y$. This quantity will be within the interval $(0,1)$ for the training data set, although it may exceed $1$ for an arbitrary data set.

For each test, we analyze the effect of the nonlinear strength of the SBS interaction, given by the $gL$ product value of the medium and the pump splitting ratio $m$ for the self-pumped nonlinear activation element, which is given by Equation \eqref{eqn:psplrat}. Here, $g$ represents the Brillouin gain and $L$ represents the interaction length. These values appear in Equation \eqref{eqn:pouttrans}. The $gL$ product decides the nonlinear shape and the parking point of the activation function, as shown in Figure \ref{fig:sbslinnonlin}. The pump splitting ratio independently changes the parking point of the activation function, which allows us to optimize it for greater performance.  It is important to note however that increasing the splitting ratio increases the power budget of the system. The parking point of the tunable activation function moves between -5 dBm to -10 dBm for the range of gL values that we have used for testing in this work, as shown in Figure 1. This means that, around the parking points, the pump power will be $100-300$ mW for a splitting ratio of 1000. Since we have allowed for a maximum output $x_k$ value of $31$ dBm, the maximum pump power required in this scenario is $3.2$ W, which is  quite a feasible power budget for the nonlinear activation elements. Better performance can generally be achieved for larger splitting ratios, which means that the best performance is restricted by the energy availability.
\begin{figure*}[!tb]
     \centering
     \begin{subfigure}[b]{0.35\textwidth}
         \includegraphics[width=\linewidth]{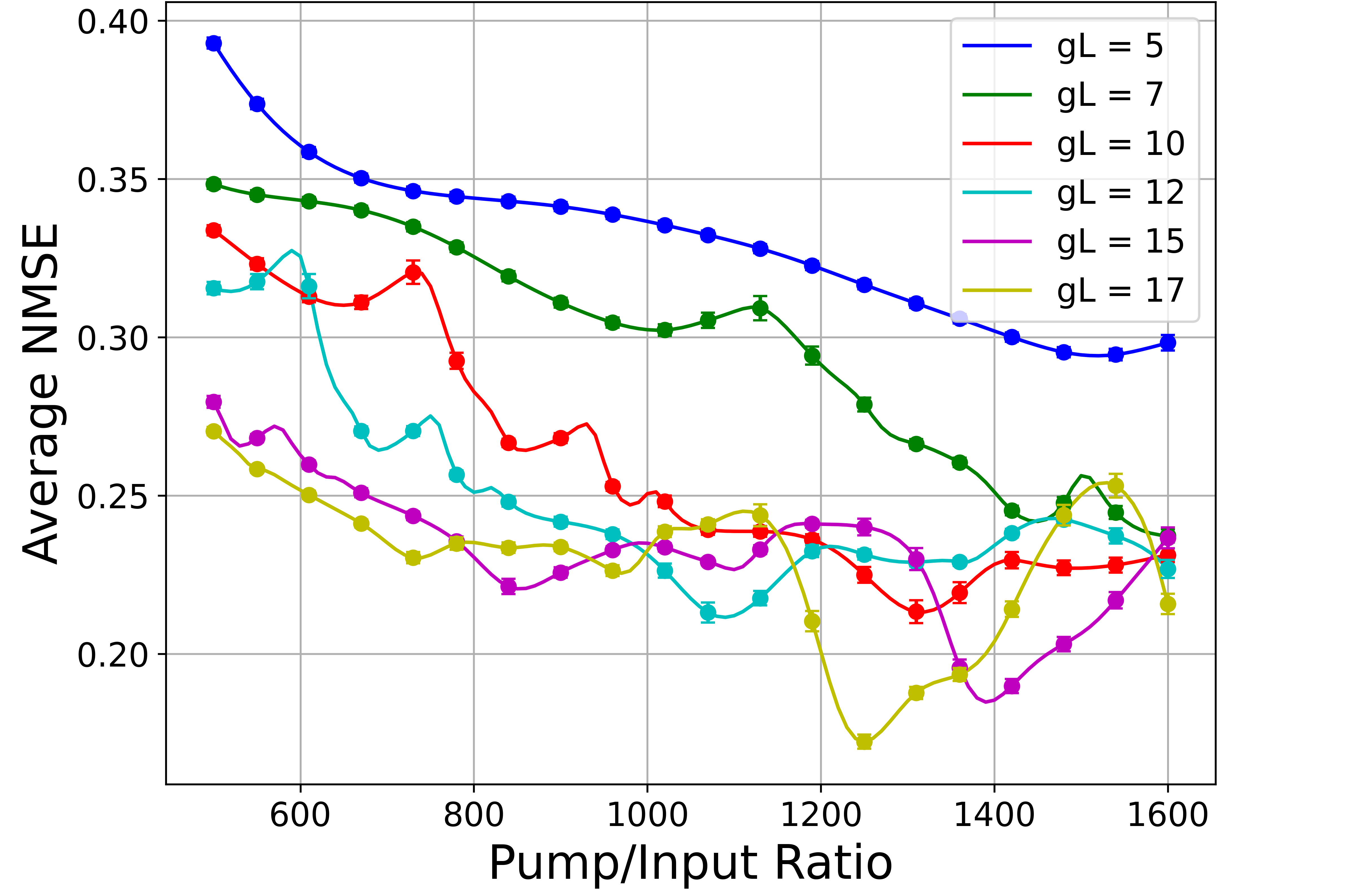}
         \caption{Average NMSE vs. Pump Splitting ratio for various $gL$, $N=10$.}
     \end{subfigure}
     \begin{subfigure}[b]{0.35\textwidth}
         \includegraphics[width=\linewidth]{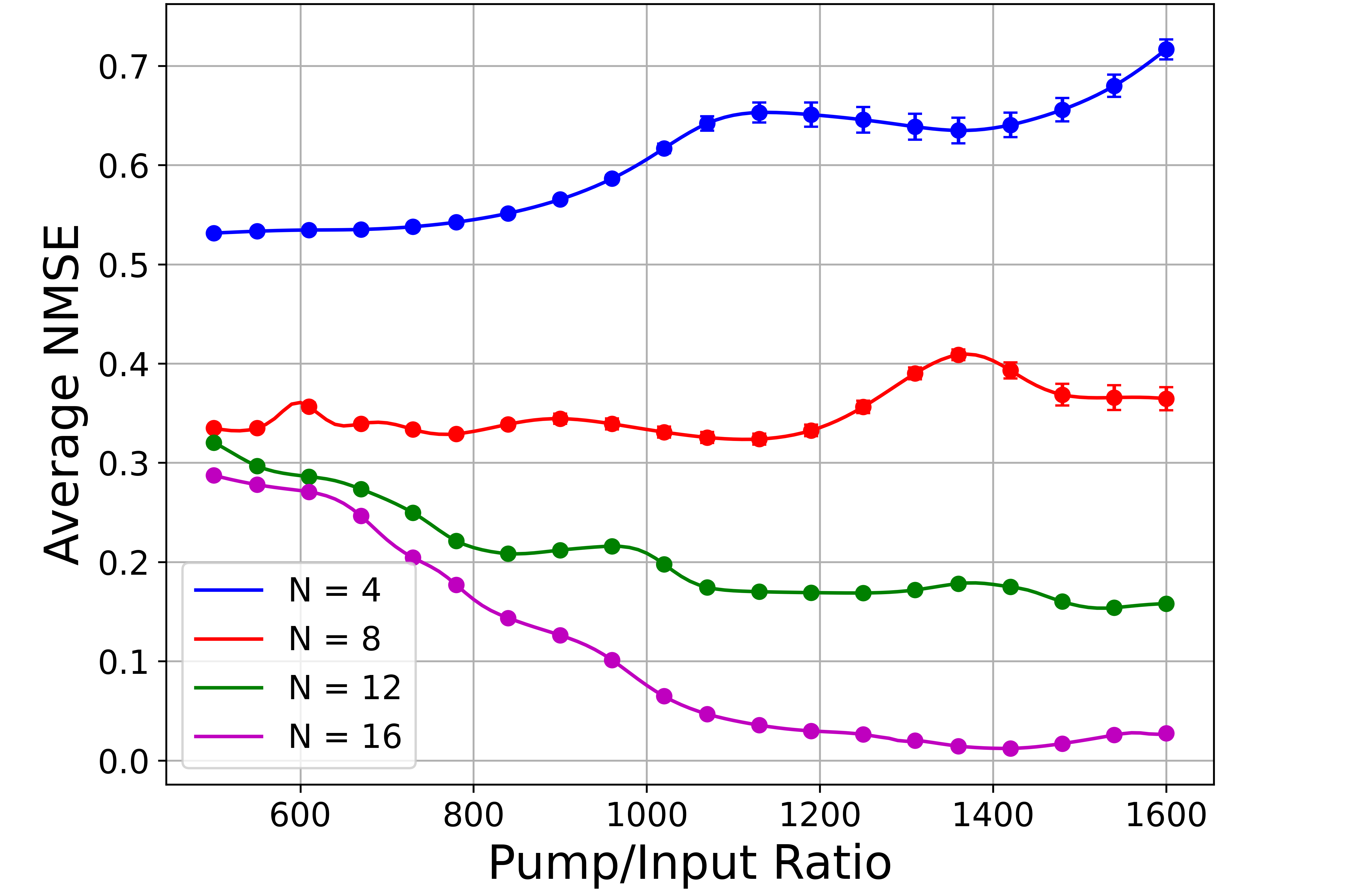}
         \caption{Average NMSE vs. Pump Splitting ratio for various $N$, $gL=12$.}
     \end{subfigure}
     \hfill
     \begin{subfigure}[b]{0.35\textwidth}
         \includegraphics[width=\linewidth]{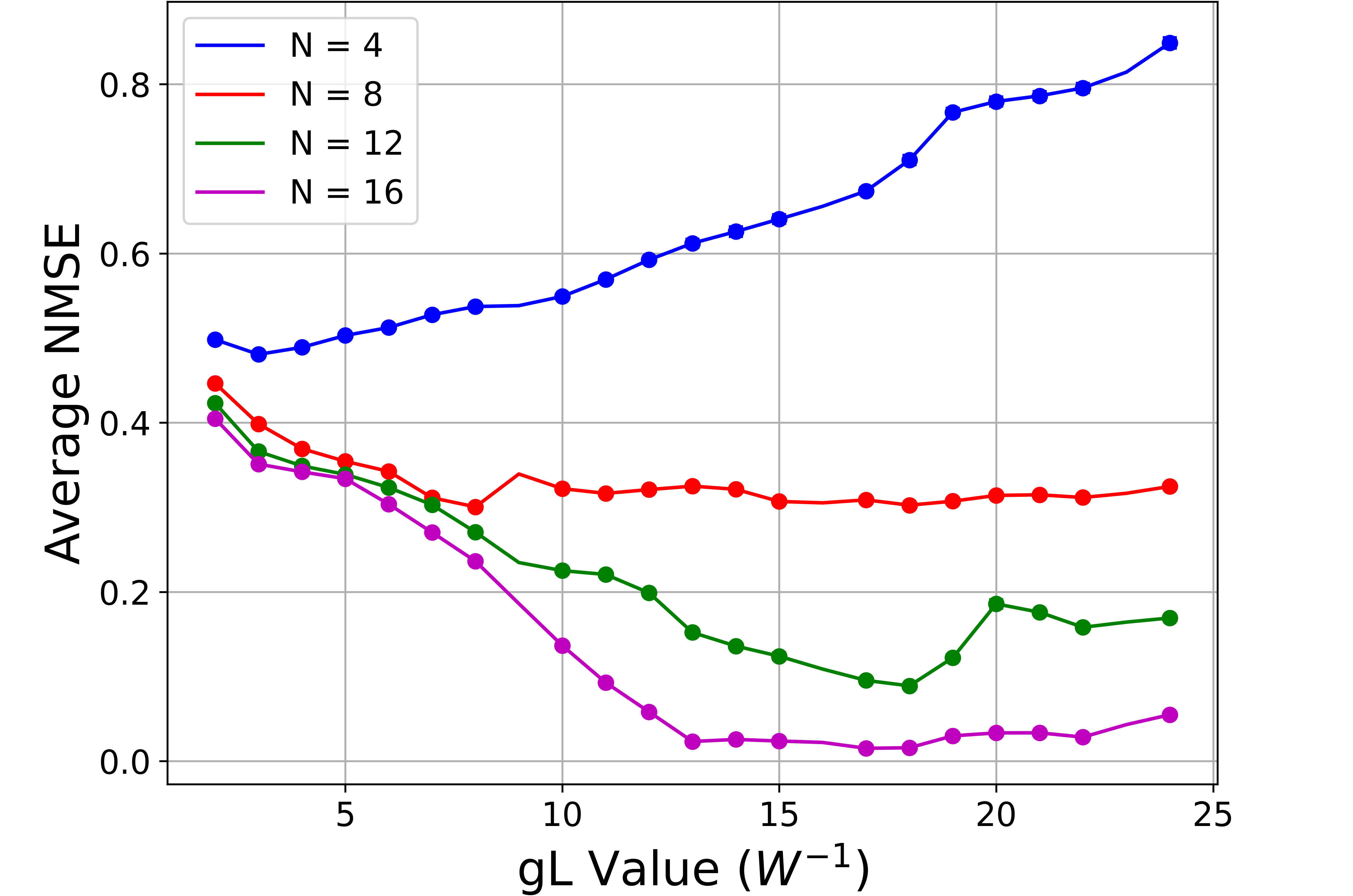}
         \caption{Average NMSE vs. $gL$ values for various $N$, $m=1000$.}
     \end{subfigure}
     \caption{NMSE vs activation function parameters for the sine square classification test.}
     \label{fig:sinsqnmse}
\end{figure*}
We also evaluate the performance as the number of nodes $N$ is increased, which shows the effect of scaling the ESN. These behaviors are task-dependent and the exact values of these parameters are subject to change with randomization of the ESN and inputs but, by extracting a range of top-performing parameters across the board, we attempt to design the most generally applicable SBS activation function and optical ESN. We do this by analyzing the behavior for 250 trials of different $A$ and $B$ matrices. These investigations are expanded on below. 

\subsection{Sine Square Classification}

This test is a time series classification task, commonly used in estimating the nonlinearity of computing reservoirs \cite{dudas2023quantum}. The input given is a series of randomly arranged sine and square waves, each 8 steps long. The task given to the ESN is to classify the current wave as sine or square. This test helps us understand the nonlinearity of the ESN. Distinguishing between smoothly varying sine waves and abruptly changing square waves requires mapping the data into a higher-dimensional hyperspace (i.e., nonlinear transformation), making classification easier. The results are shown in Figure \ref{fig:sinsq}. 

On average, we see similar performance in terms of the NMSE with the software ESN using a sigmoid activation performing marginally better. We emphasize that the comparison between the software ESN and the optical ESN is not intended to show that one approach is superior to the other. Rather, the comparison serves as a validation check for the numerical simulations of our optical ESN. We anticipate that the performance of the software ESN and the optical ESN will be quite similar, as both implementations rely on the same underlying model, differing only slightly in their nonlinear activation functions. The reported values for this instance given in the figures are 0.14 for the software ESN and 0.12 for the optical ESN. These are median values from multiple computations. 

We evaluate the effect of varying activation function parameters on the NMSE for this test. We have computed the average NMSE over 250 variations of the $A$ and $B$ matrices to see clear trends in the NMSE. From Figures \ref{fig:sinsqnmse} (a) and (b), NMSE decreases consistently with increasing pump splitting ratios, indicating improved nonlinear activation efficiency. Figure \ref{fig:sinsqnmse} (c) shows that for a fixed pump splitting ratio as well, the performance nearly saturates at lower $N$ values for $gL$ products in the range of $3-7 \; W^{-1}$, with slight ($<0.2$) improvements in NMSE for higher $gL$ values in the case of larger $N$ values. We see that for $N = 4$ nodes, the trends are difficult to see and the NMSE values are much higher. This is because the fitting is difficult for a smaller number of nodes. Therefore, we find for this task that $N = 5$ and above are required for reasonable performance, with an increase in $N$ taking the performance closer and closer to saturation near $\mathrm{NMSE} \approx 0$. We also see, from Figure \ref{fig:sinsqnmse} (a) that there is a different optimum point in the performance for each $gL$ value. The most probable reasoning for this behavior is that both the pump splitting ratio and the $gL$ value are together responsible for the optimal parking point of the activation function, resulting in local minima.
\begin{figure*}[t]
     \centering
     \begin{subfigure}[t]{0.35\textwidth}
         \includegraphics[width=\linewidth]{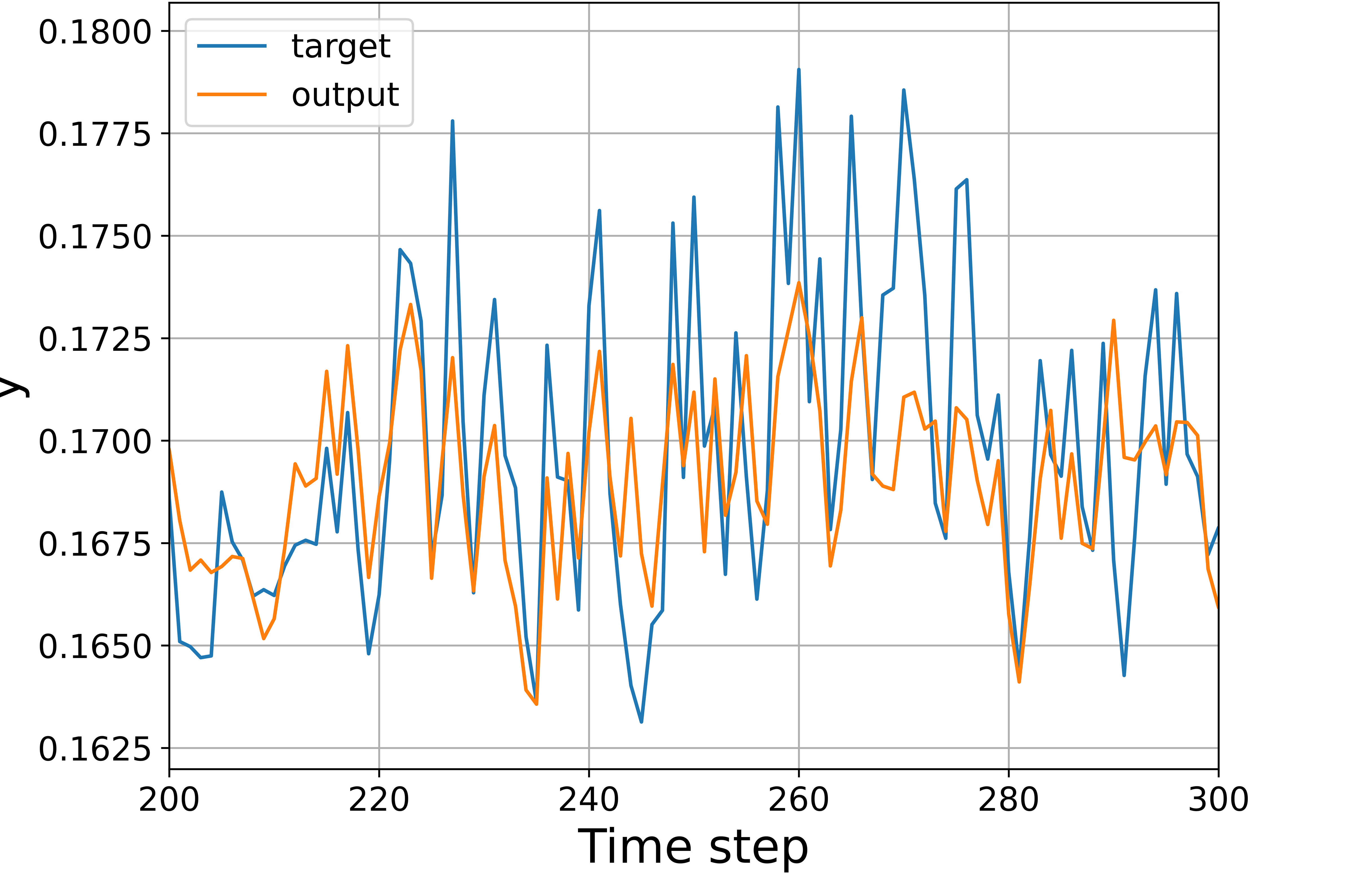}
         \caption{Software (Sigmoid) NARMA Test, NMSE = 0.57.}
     \end{subfigure}
     \begin{subfigure}[t]{0.35\textwidth}
         \includegraphics[width=\linewidth]{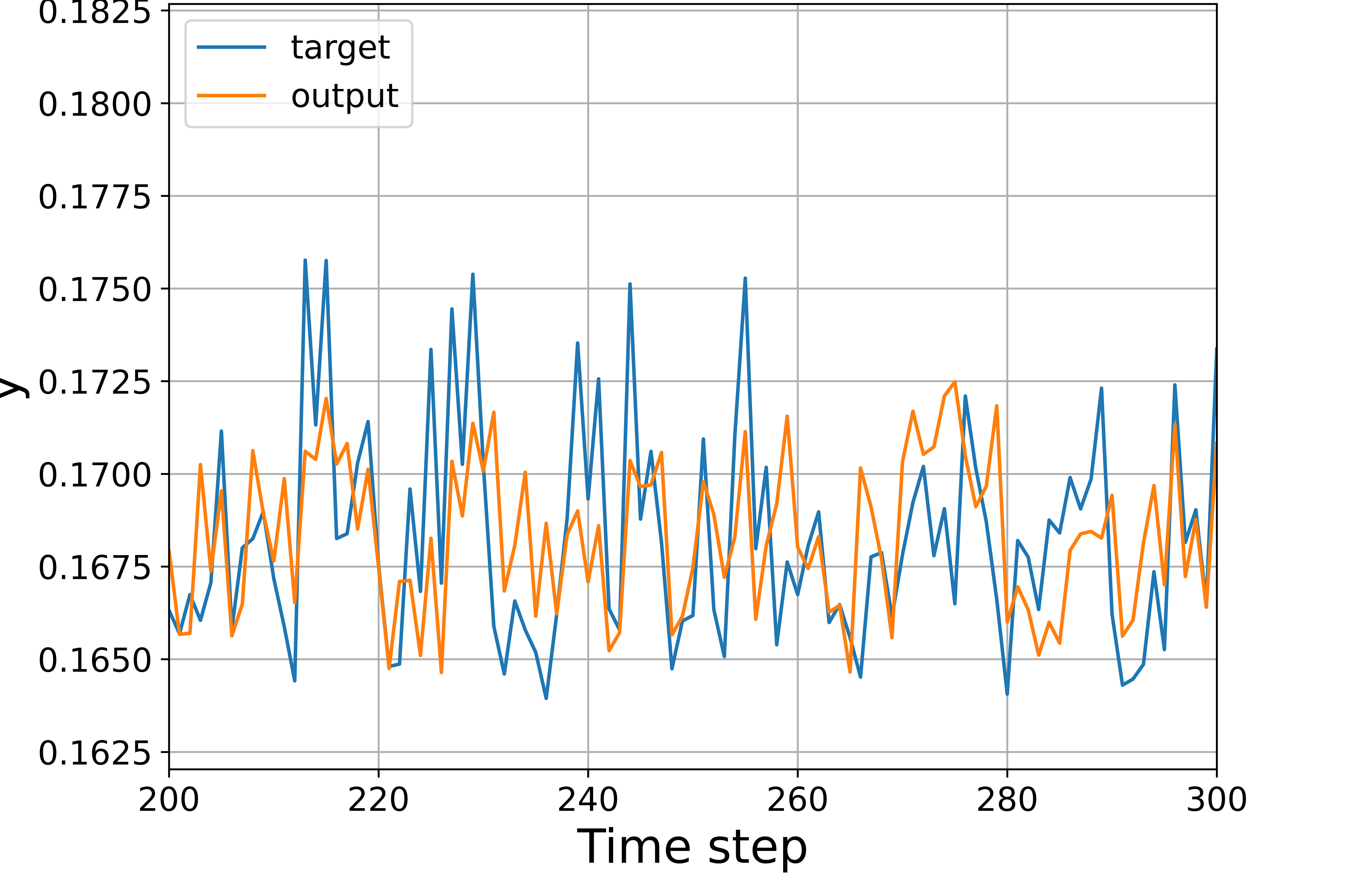}
         \caption{Optical NARMA Test, NMSE = 0.54.}
     \end{subfigure}
     \caption{Comparison of the performance of the NARMA10 test. Parameters: $N=10$, $gL=12$, $m=1500$}
     \label{fig:narmacomp}
\end{figure*}
The values reported by Slinkov et. al \cite{slinkov2024all} allow the possibility of $gL$ values from $3-7$ with fiber lengths of $3 - 6$m at reasonable threshold powers of $1.26$ W which are very practical lengths of fiber. Experimental demonstrations have shown much higher levels of nonlinearity in the right materials  \cite{eggleton2019brillouin,choudhary2016advanced}. This shows promise for the translation of this architecture into an on-chip implementation, which has the benefit of improved scalability while maintaining the other advantages of this fiber-based approach.

\subsection{NARMA10}

\begin{figure*}[!tb]
     \centering
     \begin{subfigure}[b]{0.35\textwidth}
         \includegraphics[width=\linewidth]{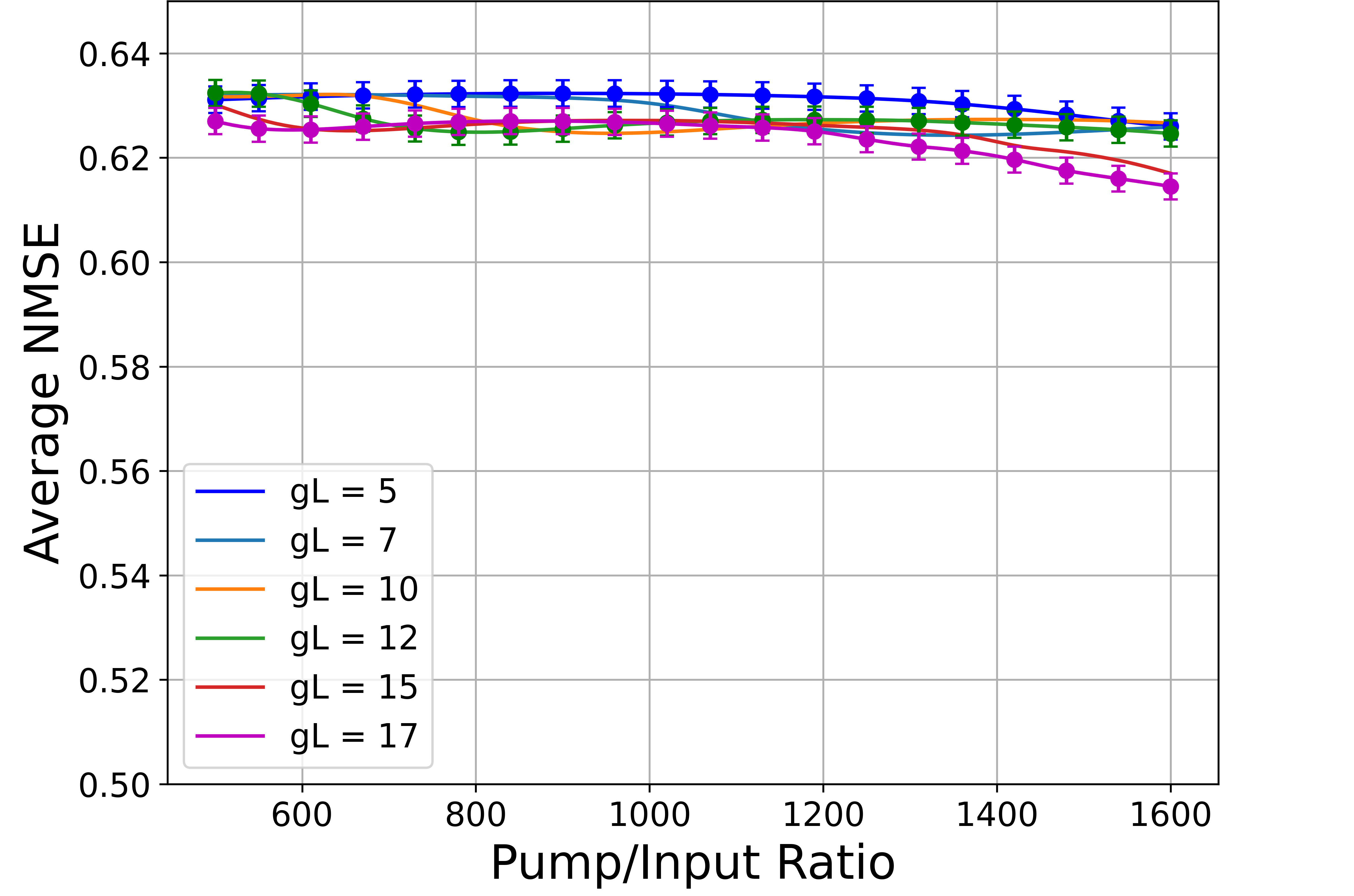}
         \caption{Average NMSE vs. Pump Splitting ratio for various $gL$, $N=10$.}
     \end{subfigure}
     \begin{subfigure}[b]{0.35\textwidth}
         \includegraphics[width=\linewidth]{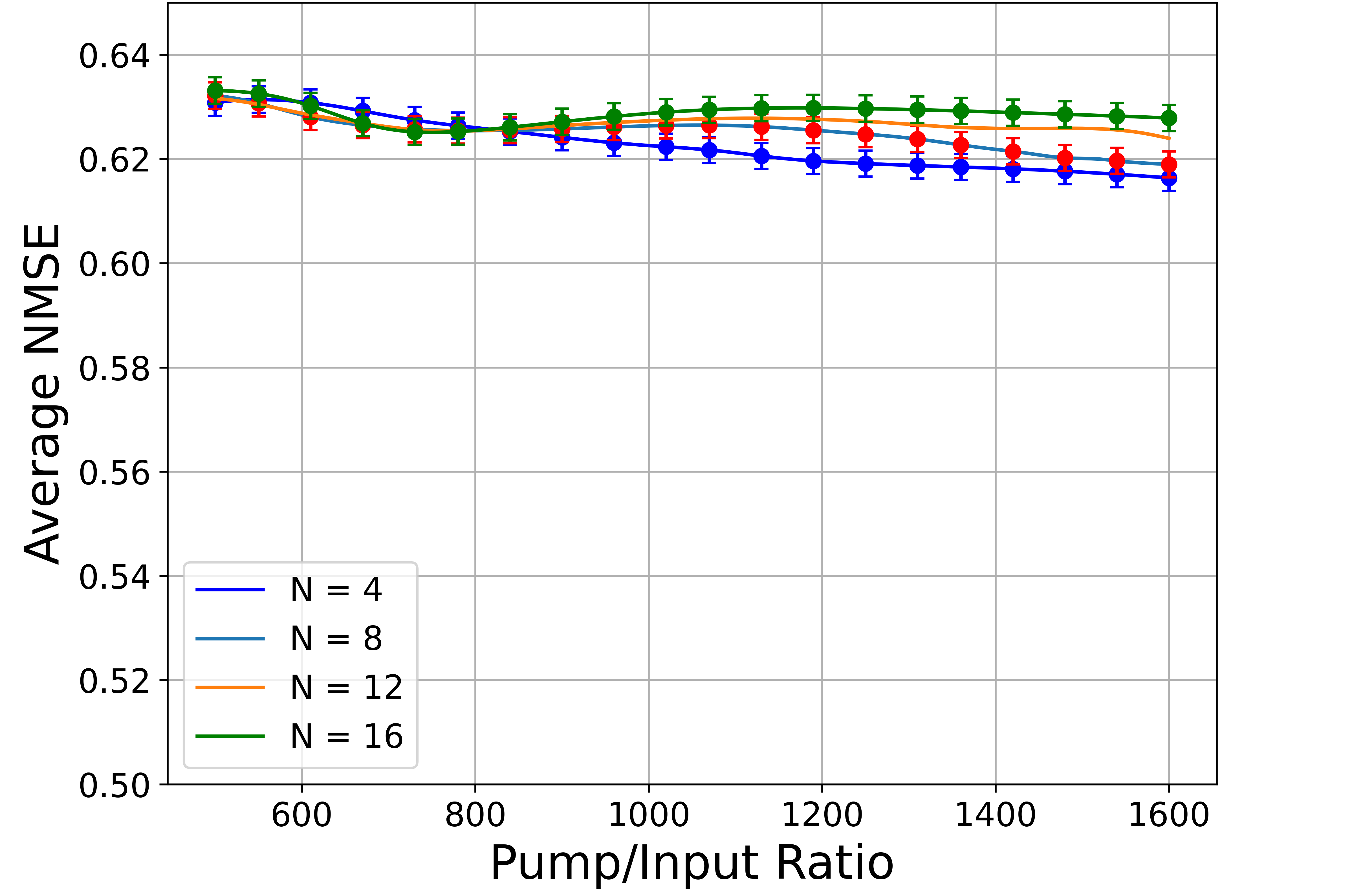}
         \caption{Average NMSE vs. Pump Splitting ratio for various $N$, $gL=12$.}
     \end{subfigure}
     \hfill
     \begin{subfigure}[b]{0.35\textwidth}
         \includegraphics[width=\linewidth]{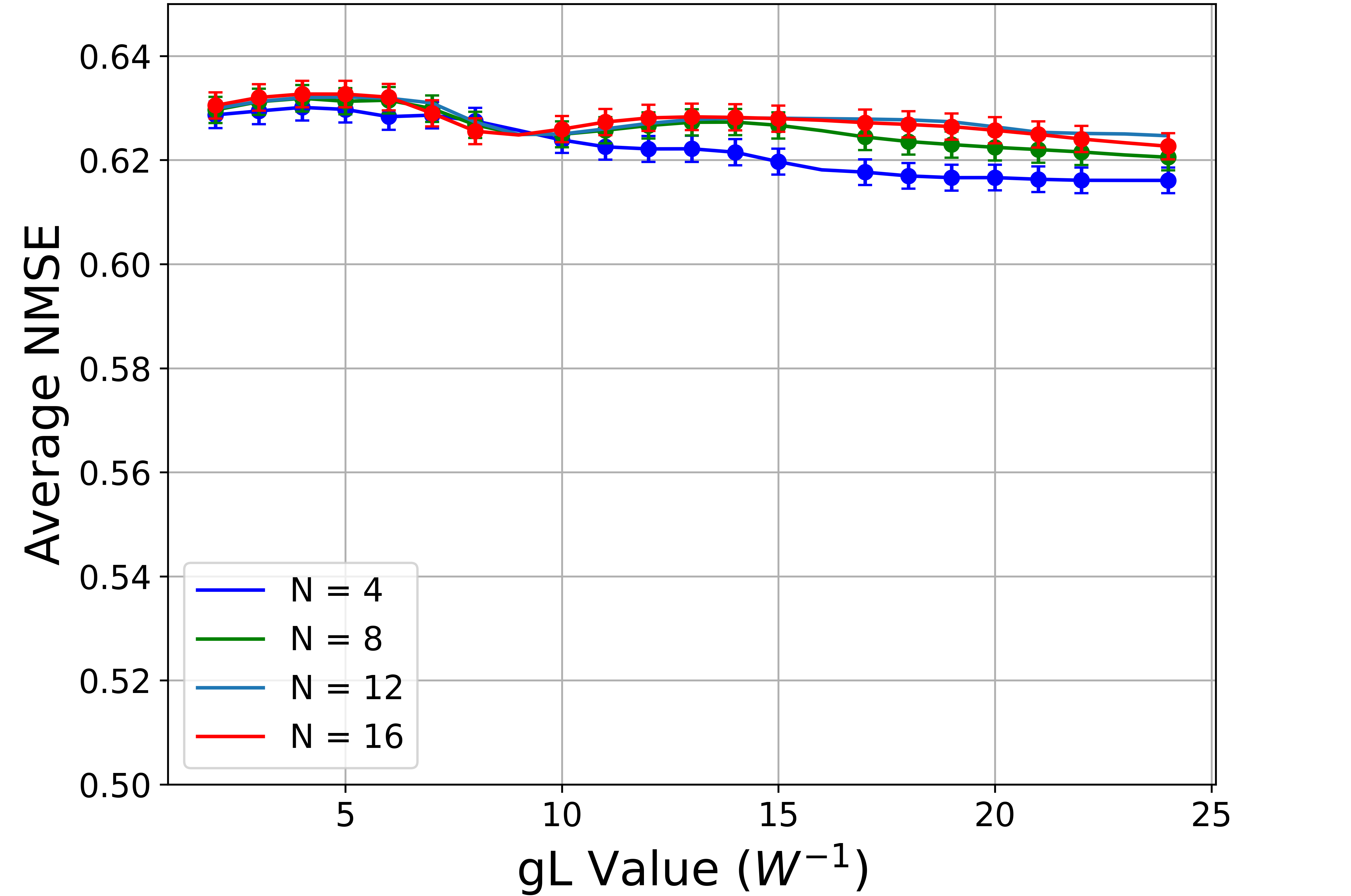}
         \caption{Average NMSE vs. $gL$ values for various $N$, $m=1000$.}
     \end{subfigure}
     \caption{NMSE vs activation function parameters for the NARMA10 test.}
     \label{fig:narmanmse}
\end{figure*}
Next, the nonlinear autoregressive moving average (NARMA) \cite{atiya2000new} task is used to test the ESN. This task mainly focuses on the polynomial algebra property of the optical ESN reservoir. The model for NARMA10, a version of NARMA popularly used for testing reservoir computers, is given by
\begin{equation}
    y_{k+1} = \alpha{}y_k + \beta{}y_k\sum_{r=0}^9 y_{k-r}+ \gamma{}u_ku_{k-9} + \delta,
    \label{eqn:narma1}
\end{equation}
\begin{equation}
    u_k = \mu + \kappa\Gamma_k,
    \label{eqn:narma2}
\end{equation}
where $y_k$ is the output at time step $k$ for input $u_k$. We have used values for the constant parameters $(\alpha, \beta, \gamma, \delta)=(0.3,0.05,1.5,0.1)$ and $\Gamma_k$ is a random variable that takes values from the interval $[-1,1]$. We have used $\mu = 0$ and $\kappa = 0.1$. This $y_k$ given by Equations \eqref{eqn:narma1} and \eqref{eqn:narma2} gives us the target sequence that we fit the ESN estimated output to. This is a tenth-order equation with a large time delay of 10 steps, making it a difficult problem. It demonstrates the capability of the reservoir computer to estimate polynomials given random inputs. The results for a software-based ESN are used to evaluate the performance of the optical ESN.

For this test, we used the following parameters of the optical ESN: $N=10, gL=12,$ and the pump splitting ratio of $m=1500$. On average we see marginally better performance from the software sigmoid ESN, but the NMSE values are comparable. We note that the performance difference between the software and optical ESNs arises primarily from differences in their nonlinear activation functions, as all other components are accurately implemented in both systems. In the instance shown in Figure \ref{fig:narmacomp}, we see an NMSE of 0.57 from the software ESN and 0.54 from the optical ESN. These are median values from multiple computations, and are quite close. This means that our optical ESN performs nearly as good as the theoretical software ideal for the same size of 10 nodes in this particular benchmark, which is a strong result.

Figure \ref{fig:narmanmse} shows the behavior of the ESN for this test with varying $gL$ and pump splitting ratio. This test is seen to prefer higher splitting ratios and higher $gL$ values, although the improvement is very slight. These results imply that a stronger pump might be better for the most general applicability of the ESN, which is corroborated by the other tests as well. The minimal variation in NMSE versus the ESN parameters when compared to the other tasks is most likely attributed to the difficulty of the NARMA10 test, which involves inputs sampled from noise. These NMSE values likely represent the saturated values achievable by the optical ESN.
\begin{figure*}[t]
     \centering
     \begin{subfigure}[b]{0.35\textwidth}
         \includegraphics[width=\linewidth]{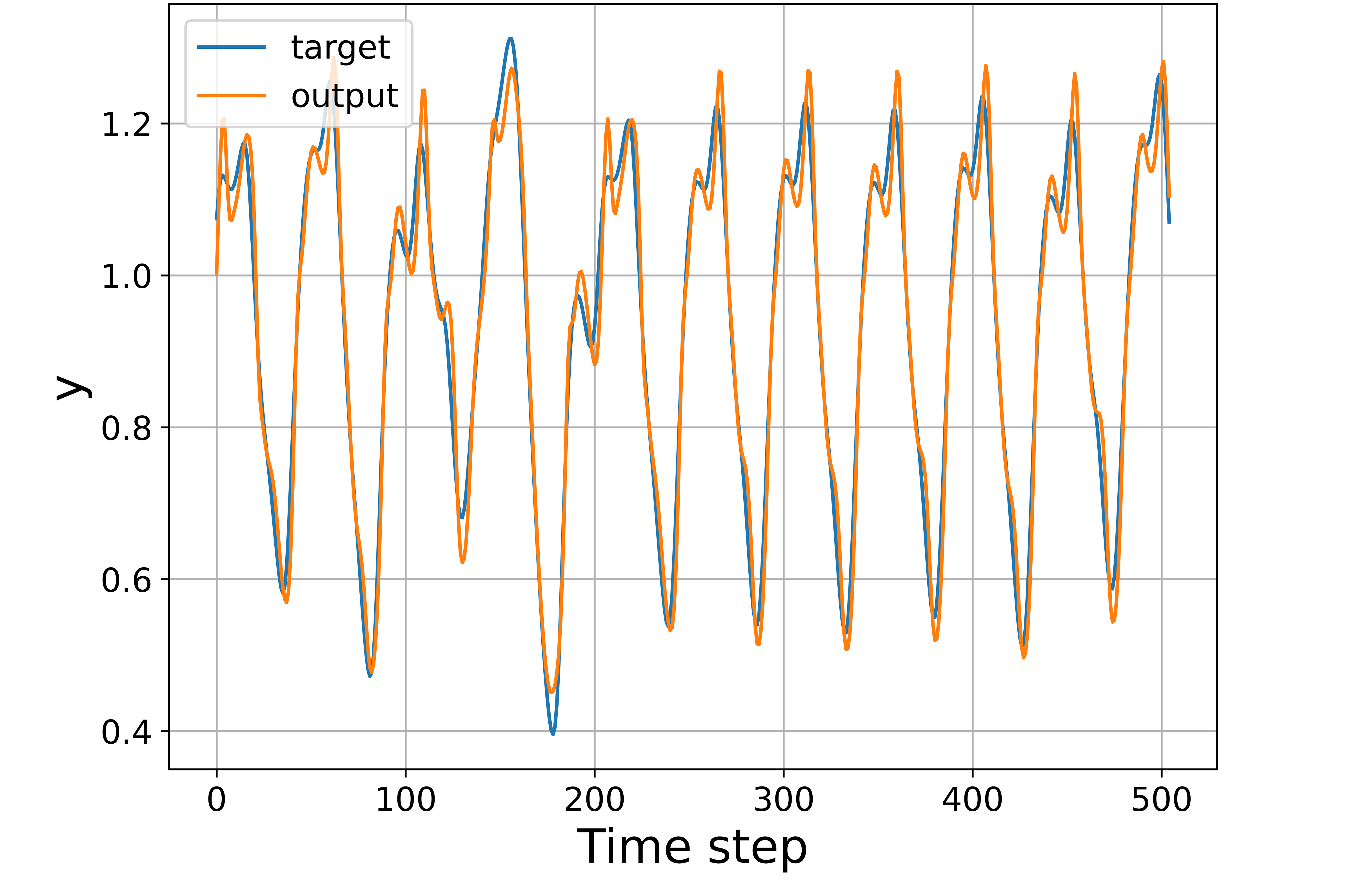}
         \caption{Software (Sigmoid) Mackey Glass Test, NMSE = 0.04.}
     \end{subfigure}
     \begin{subfigure}[b]{0.35\textwidth}
         \includegraphics[width=\linewidth]{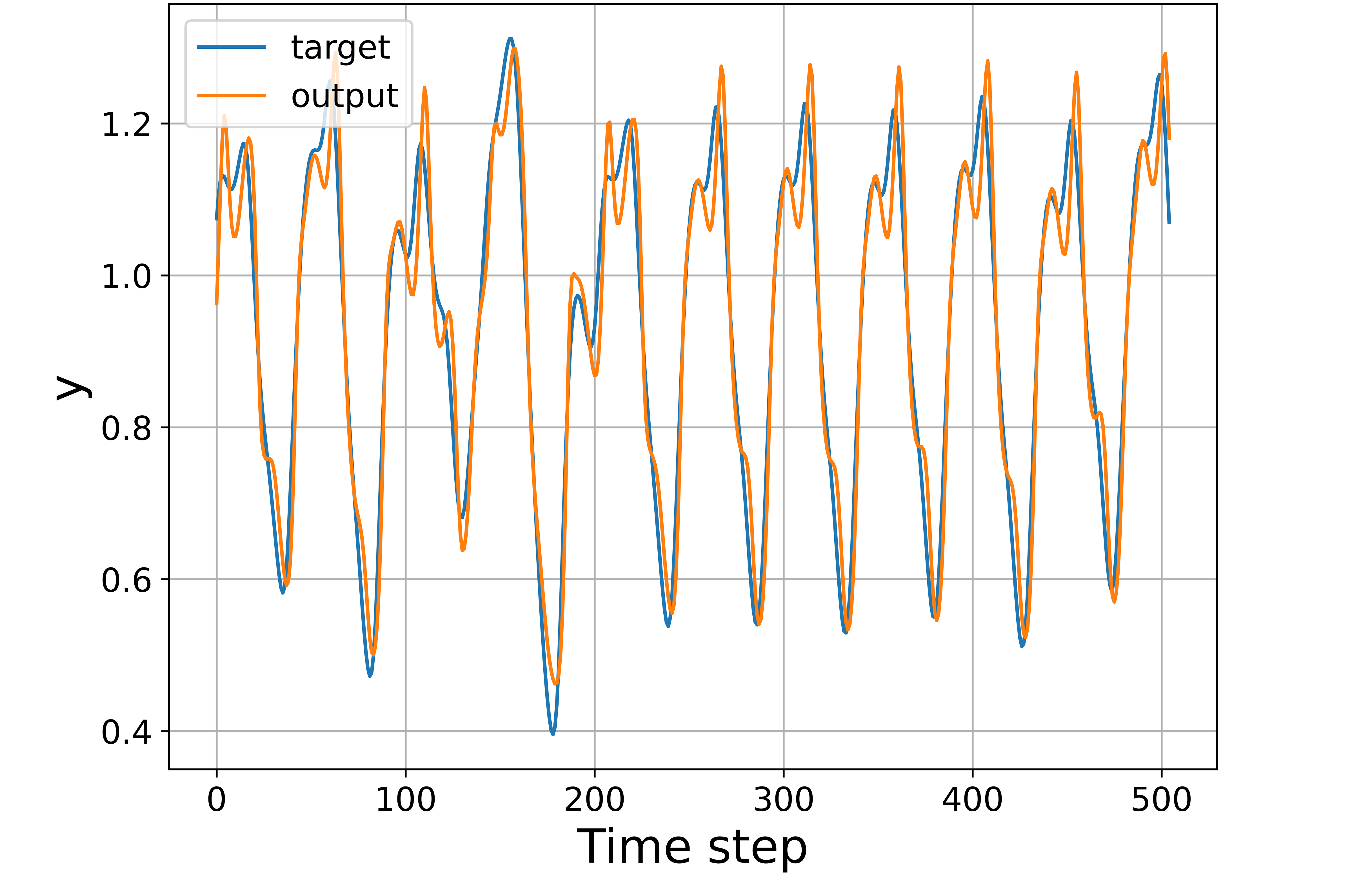}
         \caption{Optical Mackey Glass Test, NMSE = 0.06.}
     \end{subfigure}
     \caption{Comparison of the performance of the Mackey-Glass time series prediction test. Parameters: $N=10$, $gL=12$, $m=1500$}
     \label{fig:MGComp}
\end{figure*}

\subsection{Mackey-Glass}

Finally, we use the Mackey-Glass task to test the performance of the ESN against the software-based version. This test particularly focuses on the dynamic memory retention capability of the optical ESN. This test requires an approximation of the Mackey-Glass chaotic time series equation \cite{mackey1977oscillation} given by
\begin{equation}
    \frac{dy}{dt}(t) = \beta\frac{y(t-\tau)}{1 +y^n(t-\tau)} - \gamma{}y(t).
    \label{eqn:MG}
\end{equation}

The values chosen for the constants in Equation \eqref{eqn:MG} are the standard values \linebreak $(\beta,\tau,n,\gamma) = (0.2,17,10,0.1)$. The solution to this equation is numerically approximated and allowed to run for 1000 steps before being used for the task. The task given to the ESN is to predict 5 steps ahead of the input, i.e. given the input ${u_k = y_{k-5}}$ we want a successful prediction of ${y_k}$.  This benchmark evaluates the reservoir computer's memory capacity and its ability to predict chaotic temporal variations. Strong performance in this test indicates that the optical reservoir has sufficient long-term memory for effective time-based predictions, as accurately forecasting future values of a complex analytical time-series requires robust extrapolation capabilities derived from effective dynamic memory retention. We evaluate the same benchmark in the software-based sigmoid ESN, and compare the two to estimate the theoretical fit of the optical ESN. We used the following parameters for the optical ESN: $N=10, gL = 12,$ and a pump splitting ratio of $m=1500$.  
\begin{figure*}[!tb]
     \centering
     \begin{subfigure}[b]{0.35\textwidth}
         \includegraphics[width=\linewidth]{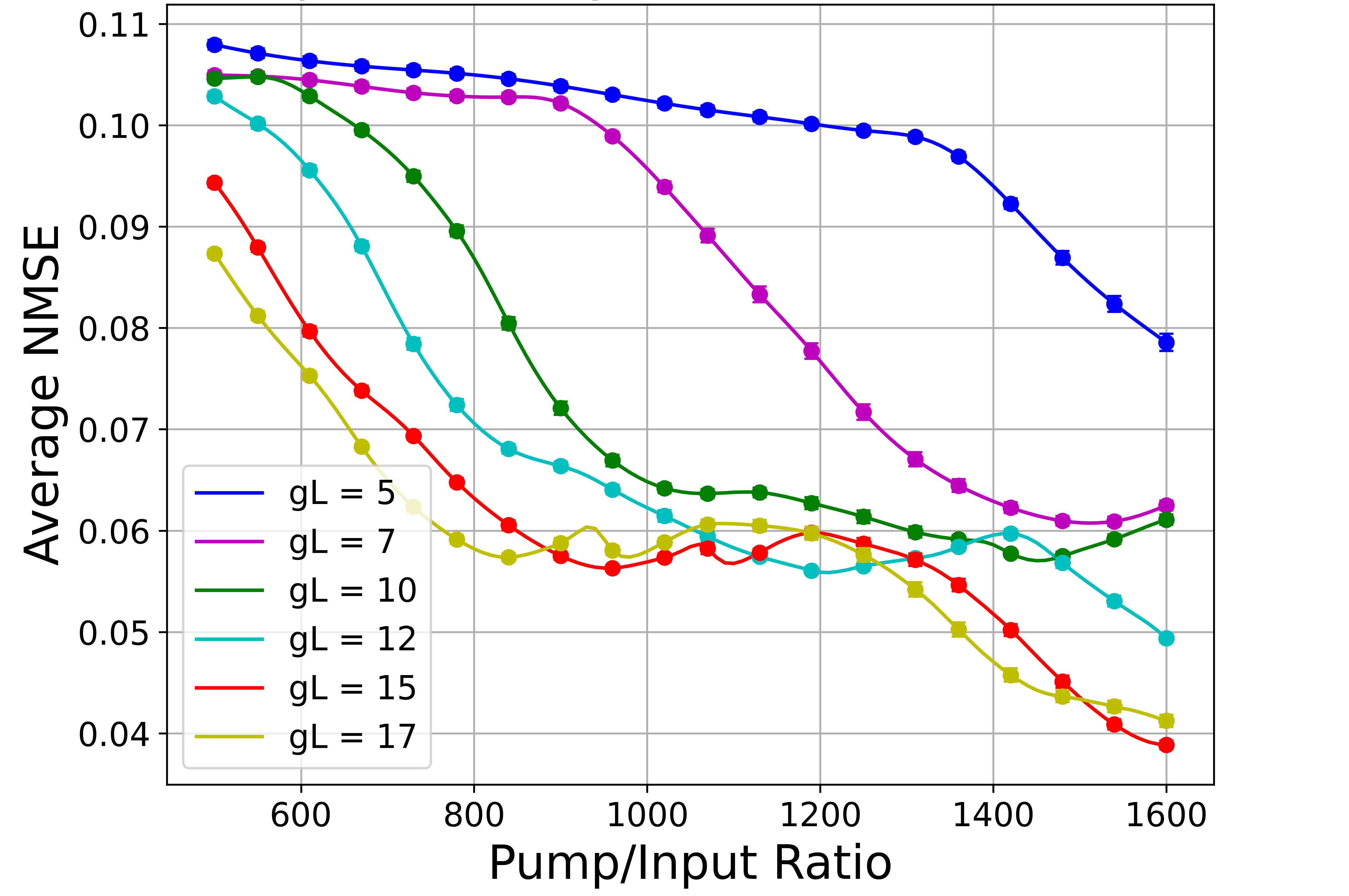}
         \caption{Average NMSE vs. Pump Splitting ratio for various $gL$, $N=10$.}
     \end{subfigure}
     \begin{subfigure}[b]{0.35\textwidth}
         \includegraphics[width=\linewidth]{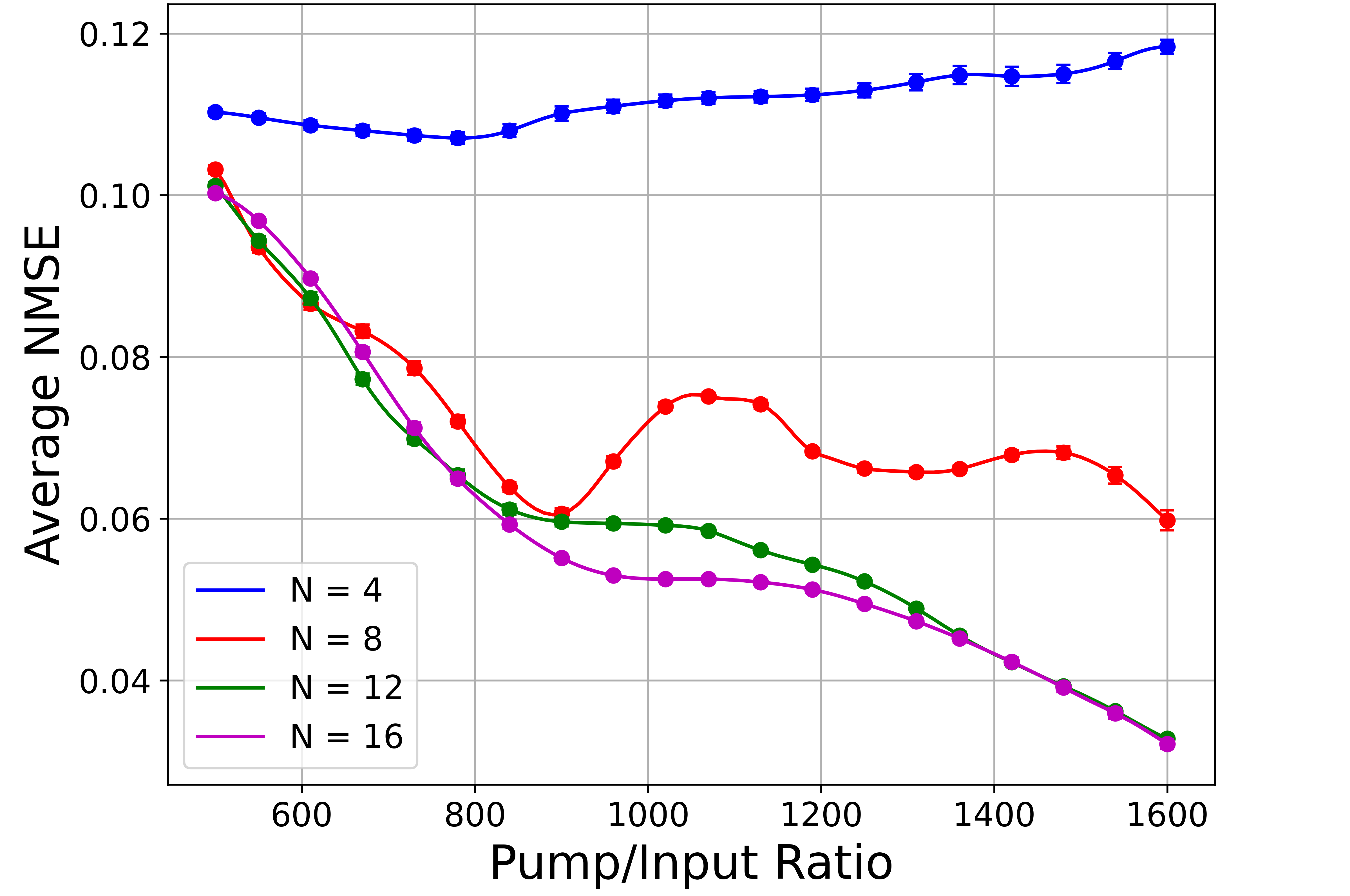}
         \caption{Average NMSE vs. Pump Splitting ratio for various $N$, $gL=12$.}
     \end{subfigure}
     \hfill
     \begin{subfigure}[b]{0.35\textwidth}
         \includegraphics[width=\linewidth]{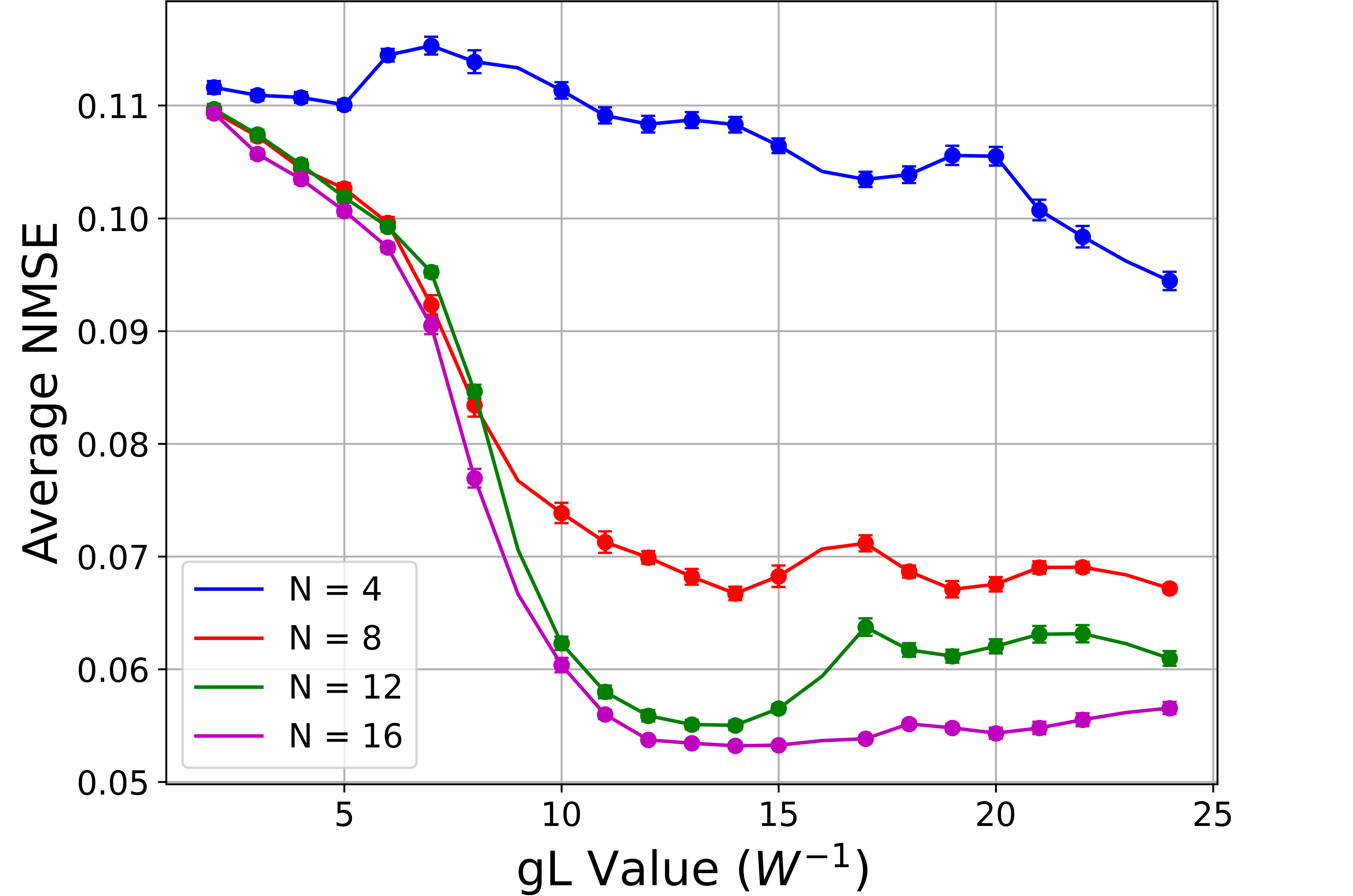}
         \caption{Average NMSE vs. $gL$ values for various $N$, $m=1000$.}
     \end{subfigure}
     \caption{NMSE vs activation function parameters for the Mackey-Glass test.}
     \label{fig:MGnmse}
\end{figure*}
For this test as well, on average we see marginally better performance from the software ESN, with the NMSE reported for the instance shown in Figure \ref{fig:MGComp} as 0.04 for the software ESN and 0.06 for the optical ESN. These NMSE values are median values seen from multiple trials. These results demonstrate that the optical ESN achieves competitive accuracy using a compact 10-node architecture, underscoring its practicality for resource-constrained applications.
\begin{figure*}[!tb]
     \centering
     \begin{subfigure}[t]{0.35\textwidth}
         \includegraphics[width=\linewidth]{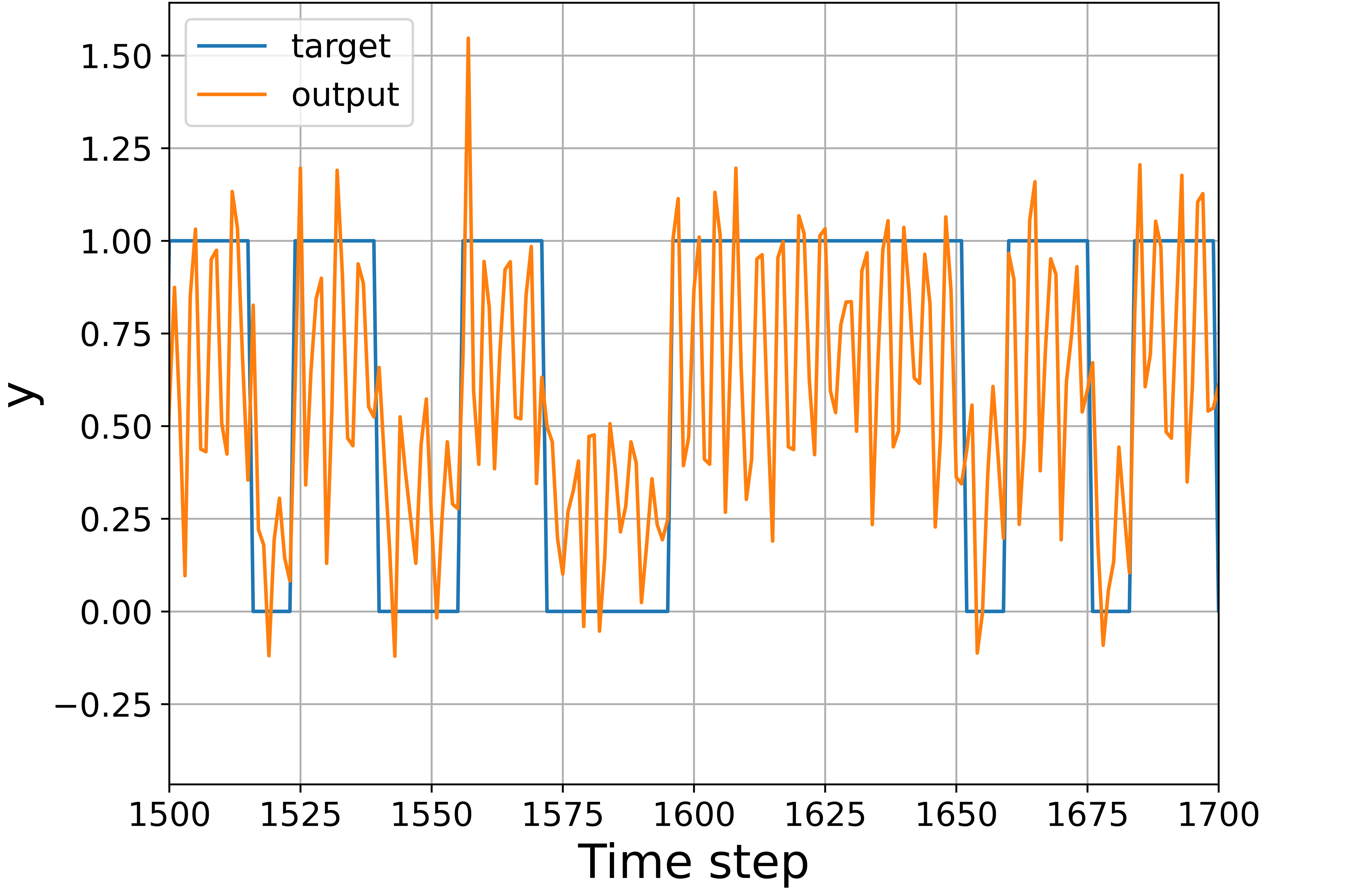}
         \caption{Optical Sine Square Classification With Noise, NMSE=0.53.}
     \end{subfigure}
     \begin{subfigure}[t]{0.35\textwidth}
         \includegraphics[width=\linewidth]{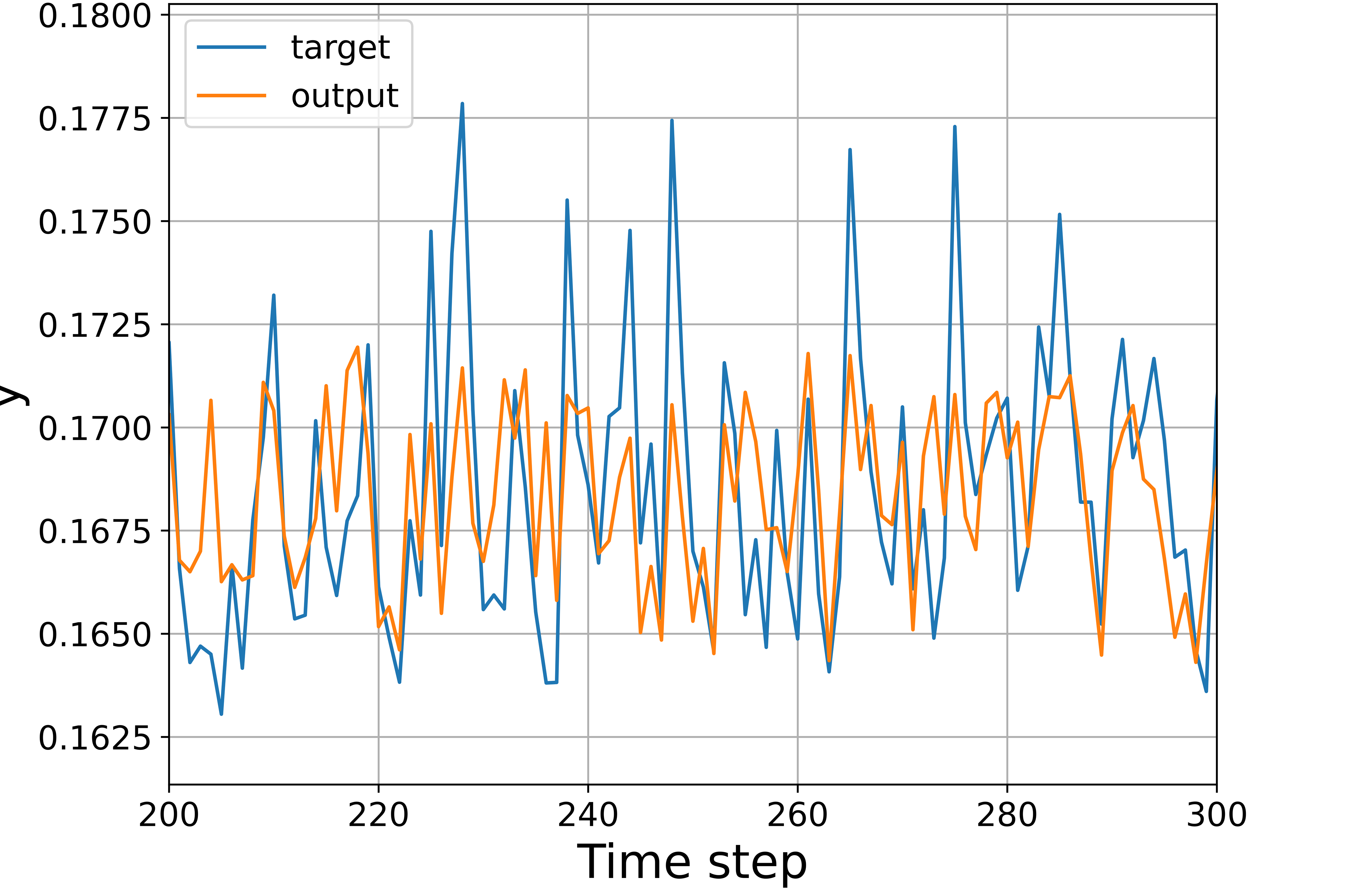}
         \caption{Optical NARMA Test With Noise, NMSE=0.75.}
     \end{subfigure}
     \hfill
     \begin{subfigure}[b]{0.35\textwidth}
         \includegraphics[width=\linewidth]{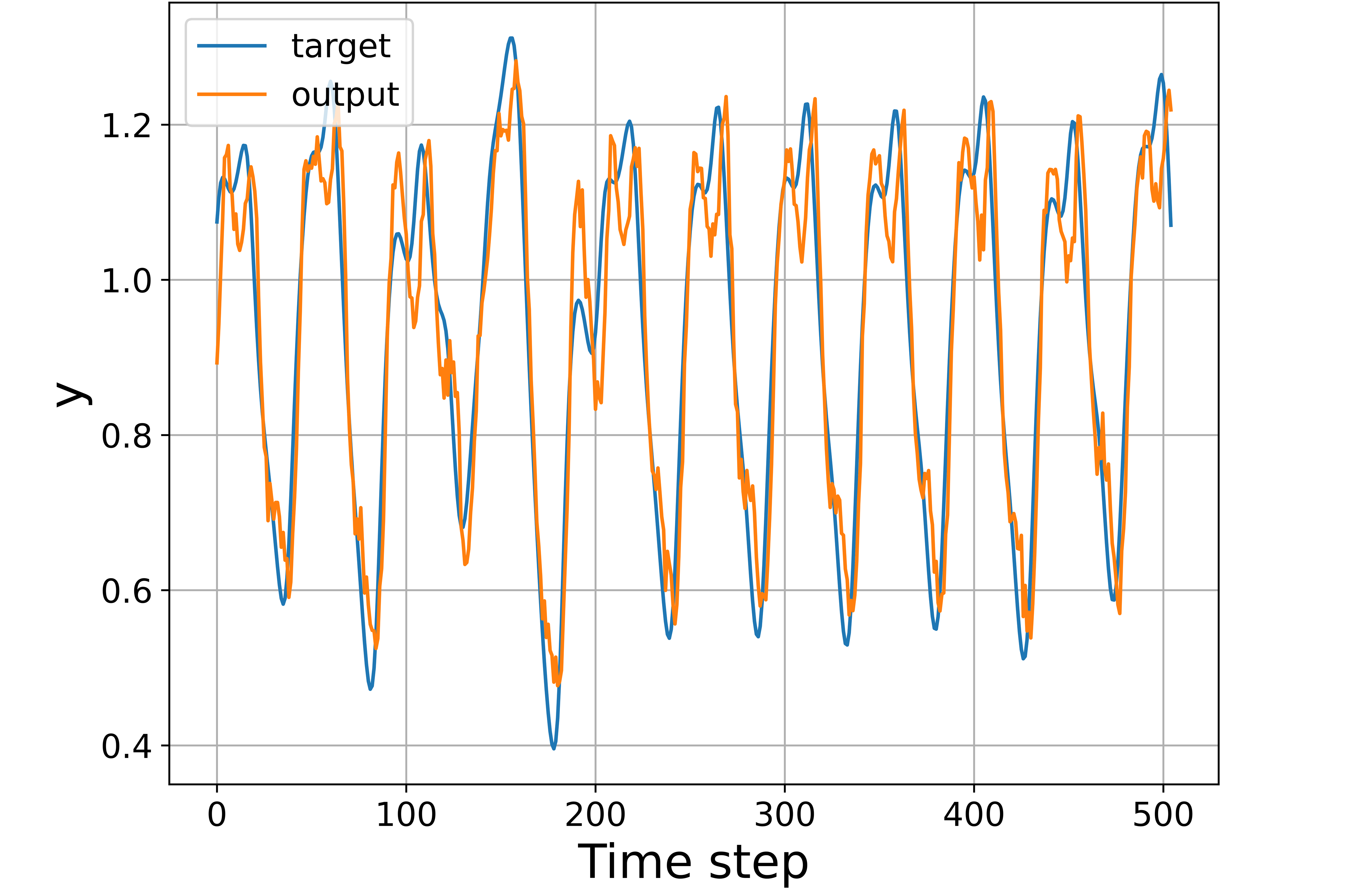}
         \caption{Optical Mackey-Glass Test With Noise, NMSE=0.15.}
     \end{subfigure}
     
     \caption{Examples of Test Performance in the Presence of Detection Noise}
     \label{fig:noise}
\end{figure*}
\begin{figure*}[!tb]
    \centering
    \begin{subfigure}[b]{0.35\textwidth}
         \includegraphics[width=\linewidth]{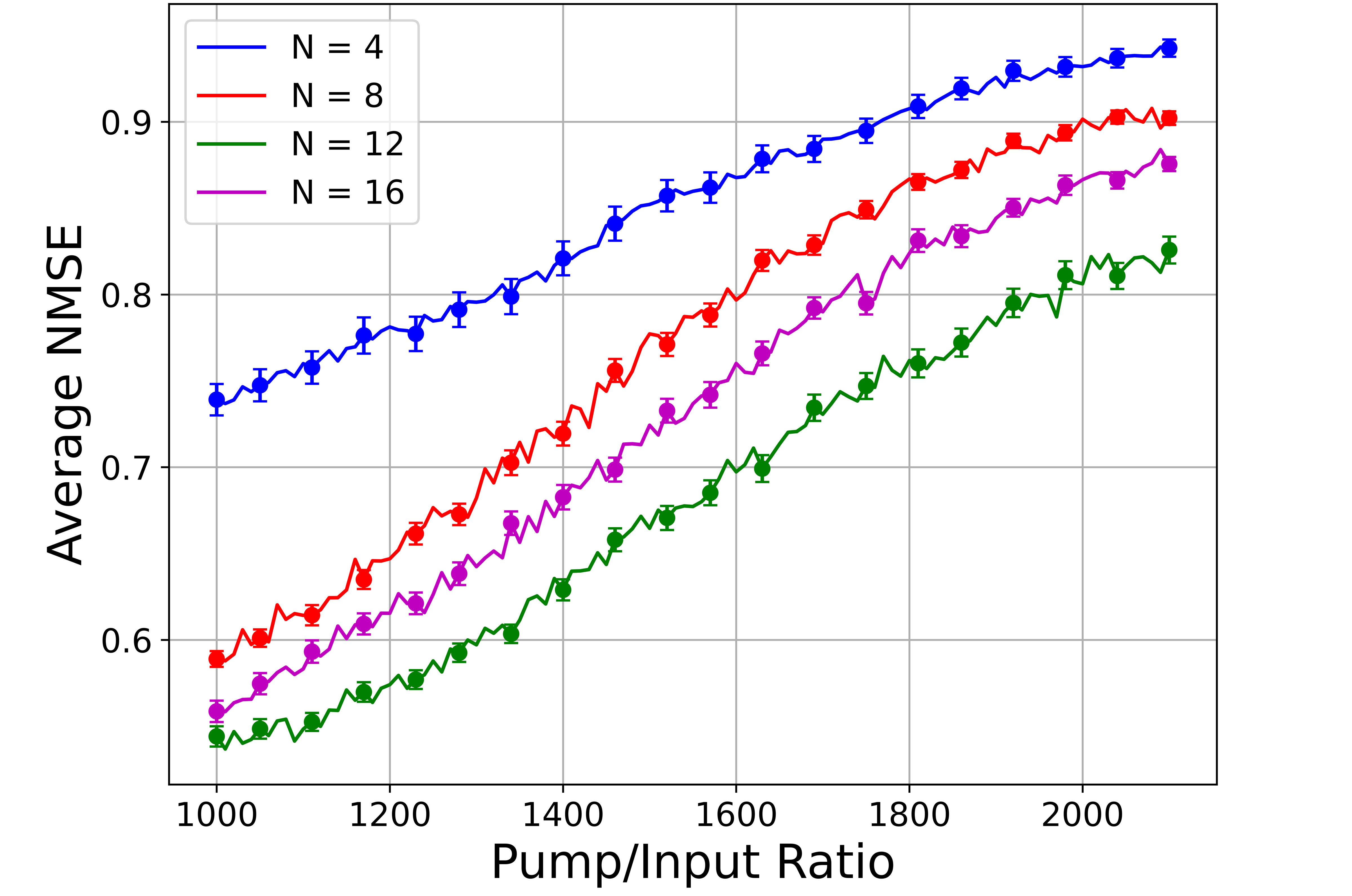}
         \caption{Average NMSE of Sine Square Classification vs. Pump Splitting ratio for various $N$, $gL=12$.}
     \end{subfigure}
     \begin{subfigure}[b]{0.35\textwidth}
         \includegraphics[width=\linewidth]{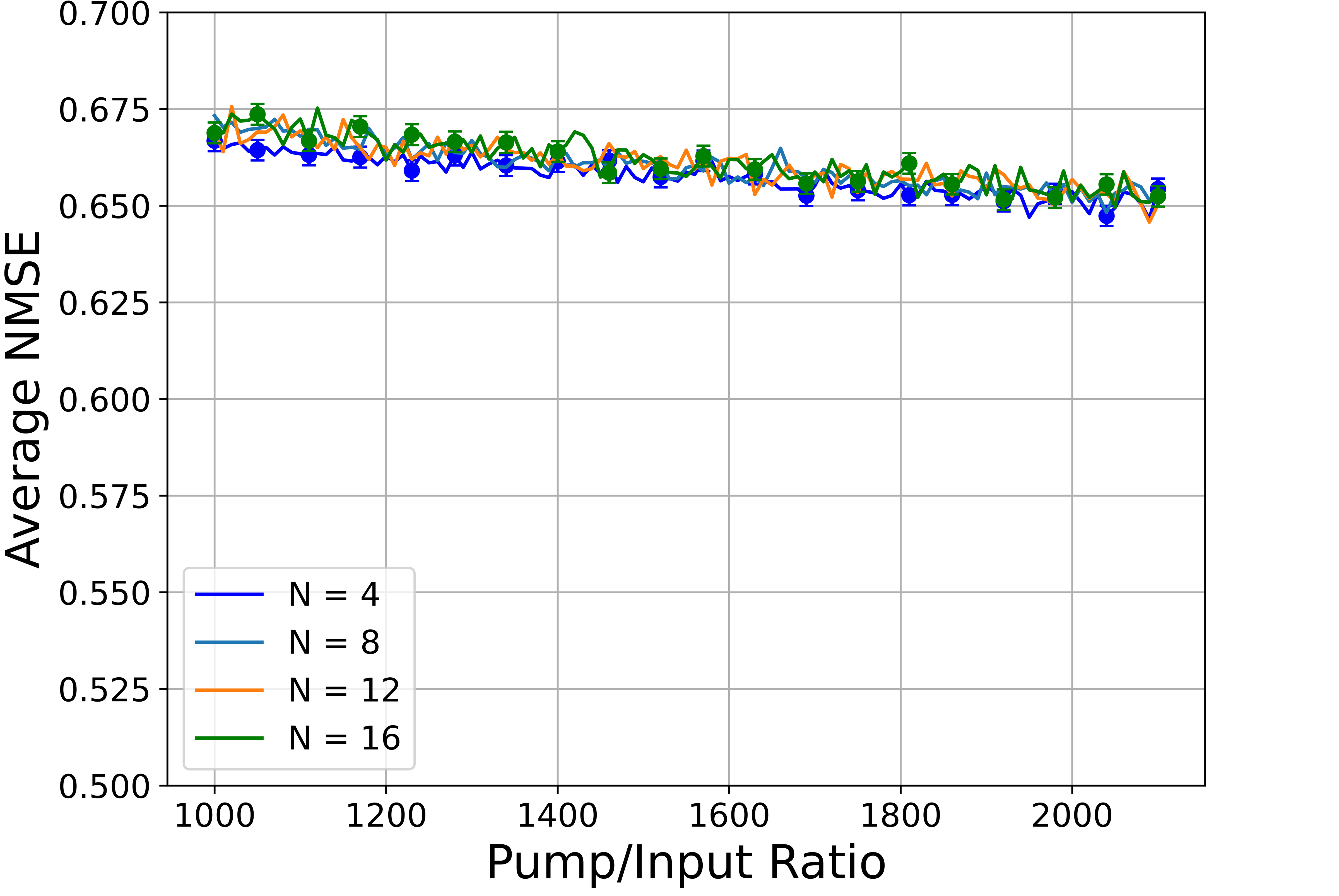}
         \caption{Average NMSE of the NARMA Test vs. Pump Splitting ratio for various $N$, $gL=12$.}
     \end{subfigure}
     \hfill
     \begin{subfigure}[b]{0.35\textwidth}
         \includegraphics[width=\linewidth]{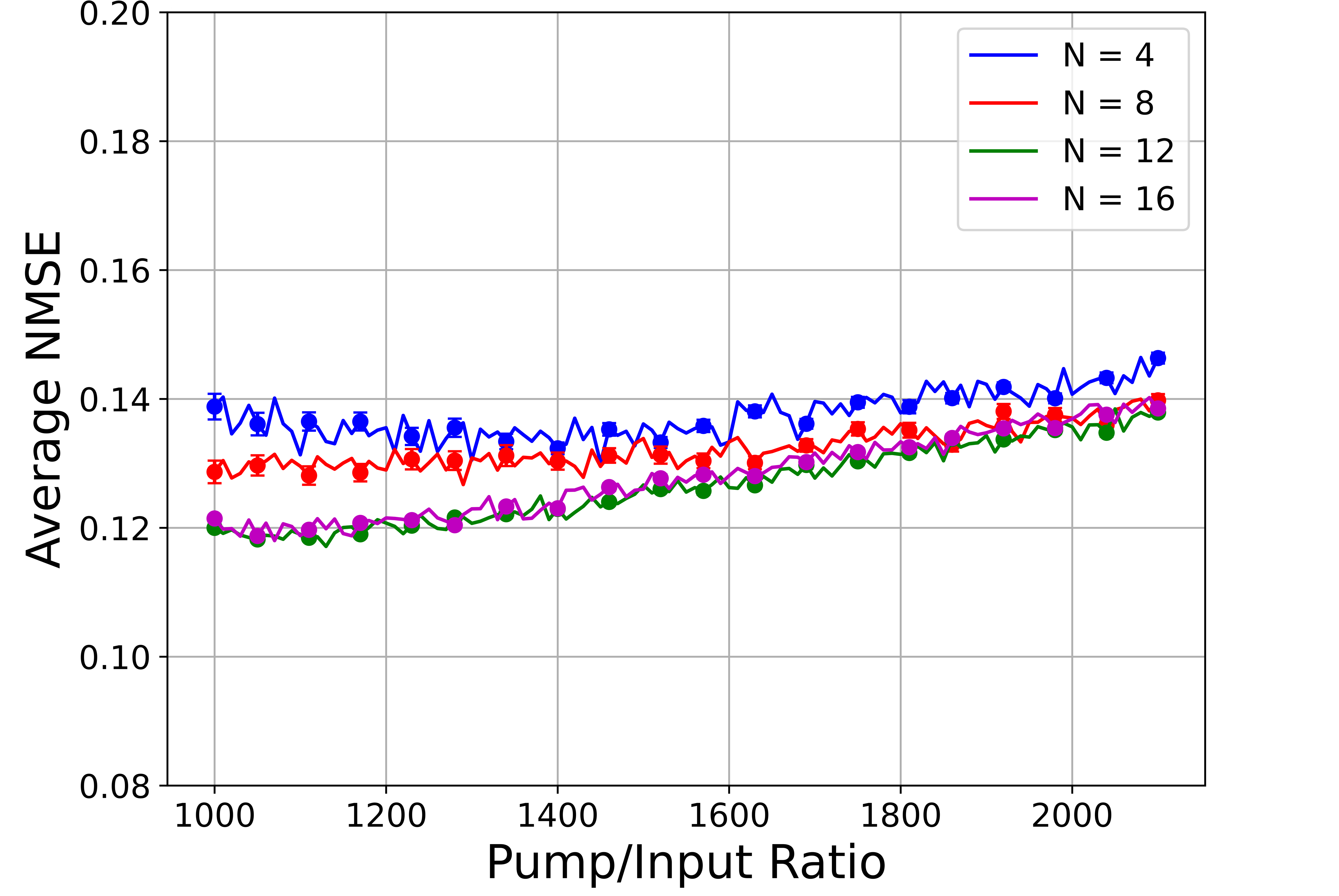}
         \caption{Average NMSE of the Mackey-Glass Test vs. Pump Splitting ratio for various $N$, $gL=12$.}
     \end{subfigure}
    \caption{Performance of Benchmark Tests in the Presence of Detection Noise}
    \label{fig:noise2}
\end{figure*}

Figure \ref{fig:MGnmse} shows the behavior with the variation of pump splitting ratio and $gL$ value. Similar to the results from the sine square classification test, the NMSE decreases with larger pump splitting ratio and shows improper fitting for $N = 4$. Although, the variation in values is slight here as well, similar to the NARMA10 test. Across all tests, it seems that a high pump splitting ratio and a $gL$ value of around $5-10 \;W^{-1}$ can ensure the best results, as the improvement from further increasing the $gL$ value is marginal at the cost of increased hardware complexity. While these behaviors can vary between tests and randomization, ensuring good performance across all three tests ensures good performance for the optical ESN in general application. In order to limit the energy consumption of the system, the pump splitting ratio can be limited to 1000. This still shows good results, as can be seen from Figures \ref{fig:sinsqnmse}, \ref{fig:narmanmse}, and \ref{fig:MGnmse}. While a larger pump splitting ratio is generally more beneficial to performance, we accept the slight degradation in exchange for enhanced feasibility of the system through a smaller pump power budget.

\subsection{Noise Performance and Negative Values}

For the case of a weak input probe undergoing high gain amplification, Brillouin amplification is known to be noisy \cite{scott1990gain}. This noise scales with pump intensity and is more prominent in cases with pump beam aberration, which can cause reduced performance by the optical reservoir \cite{nathe2023reservoir}. In our design, these intense gain values are not necessary for the matrix multiplication steps, making our reservoir relatively safe from high gain noise. To further reduce the impact, the amplified spontaneous noise (ASE) can be mitigated by the use of a band-pass filter \cite{slinkov2024all} and in our simulations, we see good performance without requiring output probe powers over 31 dBm, which would result in a negligible noise level.
 \begin{figure*}[!tb]
     \centering
     \begin{subfigure}[b]{0.35\textwidth}
         \includegraphics[width=\linewidth]{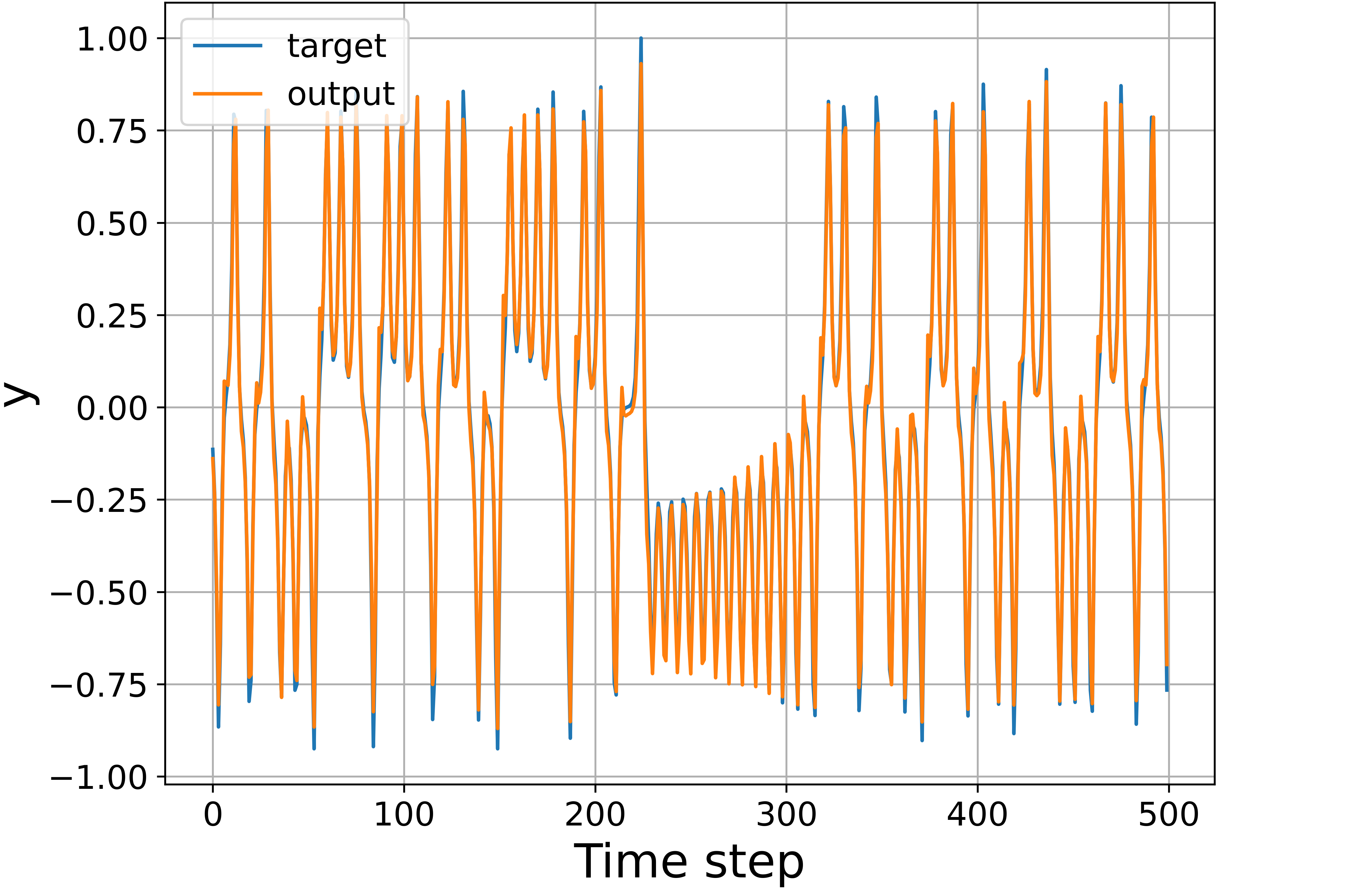}
         \caption{Optical Lorenz System Test. NMSE=0.018}
     \end{subfigure}
     \begin{subfigure}[b]{0.35\textwidth}
         \includegraphics[width=\linewidth]{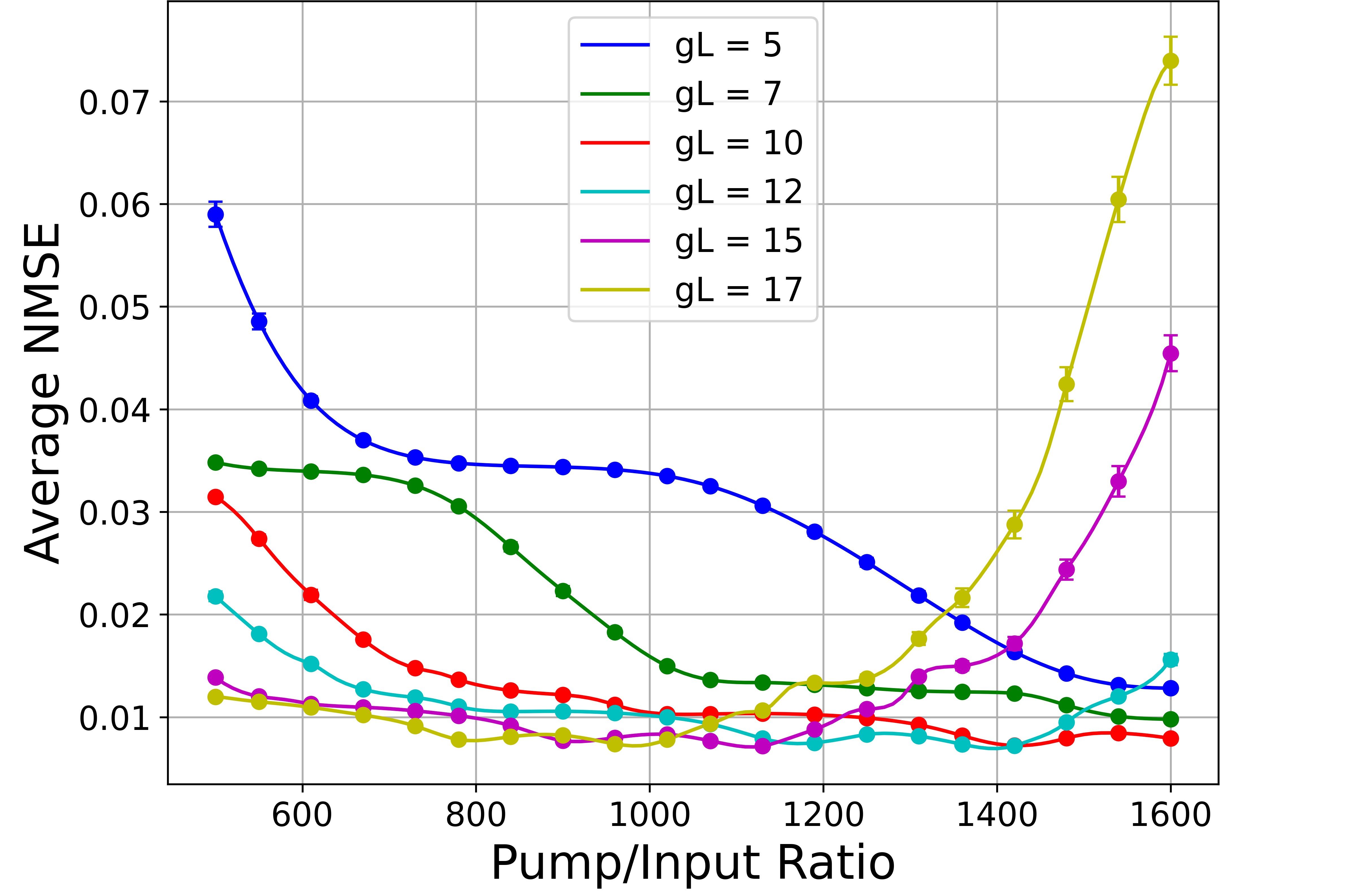}
         \caption{Variation of Lorenz System Test NMSE with Pump Splitting Ratio}
     \end{subfigure}
     \caption{Lorenz-X Prediction Results to Demonstrate Accessibility of Negative Values}
     \label{fig:lorenzcomp}
\end{figure*}
Another source of noise is optical power detection noise. This is accounted for by a simple gaussian noise model. Figure \ref{fig:noise} shows examples of the performance of the optical ESN with detector noise included, compared with Figures \ref{fig:sinsq}, \ref{fig:narmacomp}, and \ref{fig:MGComp}. For this test, we assumed the noise property of $0.32\%$ uncertainty at $1550$nm wavelength, which is a standard treatment of considering detector noise \cite{vayshenker2000optical}. In the presence of noise, the nonlinear capability is severely reduced as can be seen in the sine-square classification test results. This is mitigated by choosing an optimal  $gL$ value with a large pump splitting ratio, as detailed in the previous section. We observe from Figure \ref{fig:noise2} that the average NMSE begins to trend upward, unlike the noise free cases of Figures \ref{fig:sinsqnmse}, \ref{fig:narmanmse}, and \ref{fig:MGnmse}. The cause of this is likely that the saturation point for the specific task as a function of the pump splitting ratio $m$ is reached earlier because of the presence of detection noise.

The optical ESN shows robustness in its time-series prediction ability, even in the presence of noise. The reported median NMSE in Figure \ref{fig:noise2} for the Mackey-Glass test is 0.15. This shows that the long-term memory capability is not diminished by the presence of noise, although there is a plateau in performance with varying pump splitting ratio, as can be seen from Figure \ref{fig:noise2} (c). We can see that overall, after accounting for power detection noise in the fiber setup, there is a reduction in performance of around $0.1-0.5$ NMSE with more reduction in the classification task and less effect on prediction, implying that the efficiency of nonlinearity is affected more by the noise than the long and short term memory capabilities of the optical reservoir. These noise properties are for commercially available optical detectors. Hence, the negative impact of noise on performance can be further mitigated with the use of low noise detectors \cite{lu2018ultra}, \cite{gasmi2016fast}.

One drawback of the selected benchmark tests is that they do not cover negative values. In order to demonstrate the ability of our design to access and predict negative inputs and targets, we use the Lorenz system test. It involves predicting the x-coordinate of a chaotic time series function \cite{lorenz1963deterministic}. The results for this test are shown in Figure \ref{fig:lorenzcomp}.

We can see in Figure \ref{fig:lorenzcomp} that the architecture is able to produce negative outputs, and performs well for a task that involves negative values as well as positive values. We also see similar conclusions from the performance in this task as the other three: a gL value of $5-10 W^{-1}$ is preferred and stronger pump splitting ratios provide better performance. This shows that negative values are also accessible for this architecture by using dBm units for the optical power.

\section{Conclusion}
\label{sec:conclusion}

We have provided the design for a feasible optical ESN with minimal resource overhead and energy requirements. For the nonlinear activation and matrix multiplication, we use SBS in two regimes: the large pump regime for linear behavior, and the self-pumped regime for nonlinear behavior. In the linear regime, tuning the pump power along with required attenuation using VOAs is enough to achieve the necessary multiplications and accumulations. For the nonlinear activation step, according to reported achievable values of Brillouin gain in highly nonlinear fibers (HNLFs) \cite{slinkov2024all}, our design requires a fiber of length $3 - 6$ m to achieve the required nonlinearity for good results, which is a very practical length of fiber. Such fibers also have a reasonable threshold power to show the required nonlinear output.

Additionally, our design scales linearly with the number of nodes $N$, making it highly scalable compared to a standard matrix implementation that scales as $N^2$. It also uses only a single nonlinear activation element. However, a trade-off in computation speed has been introduced to allow for this scalable design. It is possible to achieve faster operation speeds at the cost of scalability, but this also greatly increases the hardware resource requirement, so we have decided against this approach. The nonlinear activation is optimizable by tuning the $gL$ value of the SBS amplifier as well as the pump splitting ratio $m$, which refers to the ratio of the pump to the input given to the self-pumped Brillouin amplifier. By doing so, we analyze the behavior of the optical ESN for multiple benchmark tests to find the most optimal activation function design, while ensuring low overall energy consumption.

Our matrix multiplication and accumulation steps make use of standard fiber optical components such as VOAs and fiber switches. We see comparable results to a standard software ESN using a sigmoid activation function. To show this, we chose three tasks: classification of sine and square waves, prediction of the NARMA10 sequence, and prediction of the Mackey-Glass chaotic time series. Each benchmark task demonstrates the nonlinearity, polynomial algebra capability, and long-term memory of this ESN respectively, showing that universal computations are indeed accessible with this hardware. We validated our numerical simulations of the designed optical ESNs by comparing them to a software ESN with a sigmoid activation function. The results demonstrated that both systems achieve similar performance when using an equal number of nodes. This shows the feasibility and effectiveness of an optical approach to building an ESN.

\section{Acknowledgment}

This material is based upon work supported by the U.S. Department of Energy, Office of Science, Office of Biological and Environmental Research under Award Number DE-SC0025910.

\bibliography{sample}

@article{hochreiter1997long,
  title={Long short-term memory},
  author={Hochreiter, Sepp and Schmidhuber, J{\"u}rgen},
  journal={Neural computation},
  volume={9},
  number={8},
  pages={1735--1780},
  year={1997},
  publisher={MIT press}
}

@article{grigoryeva2018echo,
  title={Echo state networks are universal},
  author={Grigoryeva, Lyudmila and Ortega, Juan-Pablo},
  journal={Neural Networks},
  volume={108},
  pages={495--508},
  year={2018},
  publisher={Elsevier}
}

@article{jaeger2001echo,
  title={The “echo state” approach to analysing and training recurrent neural networks-with an erratum note},
  author={Jaeger, Herbert},
  journal={Bonn, Germany: German National Research Center for Information Technology GMD Technical Report},
  volume={148},
  number={34},
  pages={p.13},
  year={2001},
  publisher={Bonn}
}

@article{ehlers2023improving,
  title={Improving the Performance of Echo State Networks Through Feedback},
  author={Ehlers, Peter J and Nurdin, Hendra I and Soh, Daniel},
  journal={arXiv preprint arXiv:2312.15141},
  year={2023}
}

@article{slinkov2024all,
  title={All-optical nonlinear activation function based on stimulated Brillouin scattering},
  author={Slinkov, Grigorii and Becker, Steven and Englund, Dirk and Stiller, Birgit},
  journal={arXiv preprint arXiv:2401.05135},
  year={2024}
}

@book{boyd2008nonlinear,
  title={Nonlinear optics},
  author={Boyd, Robert W and Gaeta, Alexander L and Giese, Enno},
  booktitle={Springer Handbook of Atomic, Molecular, and Optical Physics},
  pages={1097--1110},
  year={2008},
  publisher={Springer}
}

@article{zhou2022photonic,
  title={Photonic matrix multiplication lights up photonic accelerator and beyond},
  author={Zhou, Hailong and Dong, Jianji and Cheng, Junwei and Dong, Wenchan and Huang, Chaoran and Shen, Yichen and Zhang, Qiming and Gu, Min and Qian, Chao and Chen, Hongsheng and others},
  journal={Light: Science \& Applications},
  volume={11},
  number={1},
  pages={p.30},
  year={2022},
  publisher={Nature Publishing Group UK London}
}

@article{xu202111,
  title={11 TOPS photonic convolutional accelerator for optical neural networks},
  author={Xu, Xingyuan and Tan, Mengxi and Corcoran, Bill and Wu, Jiayang and Boes, Andreas and Nguyen, Thach G and Chu, Sai T and Little, Brent E and Hicks, Damien G and Morandotti, Roberto and others},
  journal={Nature},
  volume={589},
  number={7840},
  pages={44--51},
  year={2021},
  publisher={Nature Publishing Group UK London}
}

@inproceedings{yang2013chip,
  title={On-chip optical matrix-vector multiplier},
  author={Yang, Lin and Zhang, Lei and Ji, Ruiqiang},
  booktitle={Optics and Photonics for Information Processing Vii},
  volume={8855},
  pages={100--104},
  year={2013},
  organization={SPIE}
}

@article{boyd1985fading,
  title={Fading memory and the problem of approximating nonlinear operators with Volterra series},
  author={Boyd, Stephen and Chua, Leon},
  journal={IEEE Transactions on circuits and systems},
  volume={32},
  number={11},
  pages={1150--1161},
  year={1985},
  publisher={IEEE}
}

@article{wang2022optical,
  title={An optical neural network using less than 1 photon per multiplication},
  author={Wang, Tianyu and Ma, Shi-Yuan and Wright, Logan G and Onodera, Tatsuhiro and Richard, Brian C and McMahon, Peter L},
  journal={Nature Communications},
  volume={13},
  number={1},
  pages={p.123},
  year={2022},
  publisher={Nature Publishing Group UK London}
}

@article{lorenz1963deterministic,
  title={Deterministic nonperiodic flow},
  author={Lorenz, Edward N},
  journal={Journal of atmospheric sciences},
  volume={20},
  number={2},
  pages={130--141},
  year={1963}
}

@article{eggleton2019brillouin,
  title={Brillouin integrated photonics},
  author={Eggleton, Benjamin J and Poulton, Christopher G and Rakich, Peter T and Steel, Michael J and Bahl, Gaurav},
  journal={Nature Photonics},
  volume={13},
  number={10},
  pages={664--677},
  year={2019},
  publisher={Nature Publishing Group UK London}
}

@article{choudhary2016advanced,
  title={Advanced integrated microwave signal processing with giant on-chip Brillouin gain},
  author={Choudhary, Amol and Morrison, Blair and Aryanfar, Iman and Shahnia, Shayan and Pagani, Mattia and Liu, Yang and Vu, Khu and Madden, Stephen and Marpaung, David and Eggleton, Benjamin J},
  journal={Journal of lightwave technology},
  volume={35},
  number={4},
  pages={846--854},
  year={2016},
  publisher={IEEE}
}

@article{atiya2000new,
  title={New results on recurrent network training: unifying the algorithms and accelerating convergence},
  author={Atiya, Amir F and Parlos, Alexander G},
  journal={IEEE transactions on neural networks},
  volume={11},
  number={3},
  pages={697--709},
  year={2000},
  publisher={IEEE}
}

@article{mackey1977oscillation,
  title={Oscillation and chaos in physiological control systems},
  author={Mackey, Michael C and Glass, Leon},
  journal={Science},
  volume={197},
  number={4300},
  pages={287--289},
  year={1977},
  publisher={American Association for the Advancement of Science}
}

@article{yan2024emerging,
  title={Emerging opportunities and challenges for the future of reservoir computing},
  author={Yan, Min and Huang, Can and Bienstman, Peter and Tino, Peter and Lin, Wei and Sun, Jie},
  journal={Nature Communications},
  volume={15},
  number={1},
  pages={p.2056},
  year={2024},
  publisher={Nature Publishing Group UK London}
}

@article{shastri2021photonics,
  title={Photonics for artificial intelligence and neuromorphic computing},
  author={Shastri, Bhavin J and Tait, Alexander N and Ferreira de Lima, Thomas and Pernice, Wolfram HP and Bhaskaran, Harish and Wright, C David and Prucnal, Paul R},
  journal={Nature Photonics},
  volume={15},
  number={2},
  pages={102--114},
  year={2021},
  publisher={Nature Publishing Group UK London}
}

@article{duport2012all,
  title={All-optical reservoir computing},
  author={Duport, Fran{\c{c}}ois and Schneider, Bendix and Smerieri, Anteo and Haelterman, Marc and Massar, Serge},
  journal={Optics express},
  volume={20},
  number={20},
  pages={22783--22795},
  year={2012},
  publisher={Optica Publishing Group}
}

@article{phang2023photonic,
  title={Photonic reservoir computing enabled by stimulated Brillouin scattering},
  author={Phang, Sendy},
  journal={Optics Express},
  volume={31},
  number={13},
  pages={22061--22074},
  year={2023},
  publisher={Optica Publishing Group}
}

@article{stone1948generalized,
  title={The generalized Weierstrass approximation theorem},
  author={Stone, Marshall H},
  journal={Mathematics Magazine},
  volume={21},
  number={5},
  pages={237--254},
  year={1948},
  publisher={JSTOR}
}

@article{gauthier2021next,
  title={Next generation reservoir computing},
  author={Gauthier, Daniel J and Bollt, Erik and Griffith, Aaron and Barbosa, Wendson AS},
  journal={Nature communications},
  volume={12},
  number={1},
  pages={1--8},
  year={2021},
  publisher={Nature Publishing Group}
}

@article{pathak2018model,
  title={Model-free prediction of large spatiotemporally chaotic systems from data: A reservoir computing approach},
  author={Pathak, Jaideep and Hunt, Brian and Girvan, Michelle and Lu, Zhixin and Ott, Edward},
  journal={Physical review letters},
  volume={120},
  number={2},
  pages={p.024102},
  year={2018},
  publisher={APS}
}

@article{vlachas2020backpropagation,
  title={Backpropagation algorithms and reservoir computing in recurrent neural networks for the forecasting of complex spatiotemporal dynamics},
  author={Vlachas, Pantelis-Rafail and Pathak, Jaideep and Hunt, Brian R and Sapsis, Themistoklis P and Girvan, Michelle and Ott, Edward and Koumoutsakos, Petros},
  journal={Neural Networks},
  volume={126},
  pages={191--217},
  year={2020},
  publisher={Elsevier}
}

@article{zimmermann2018observing,
  title={Observing spatio-temporal dynamics of excitable media using reservoir computing},
  author={Zimmermann, Roland S and Parlitz, Ulrich},
  journal={Chaos: An Interdisciplinary Journal of Nonlinear Science},
  volume={28},
  number={4},
  year={2018},
  publisher={AIP Publishing}
}

@article{canaday2018rapid,
  title={Rapid time series prediction with a hardware-based reservoir computer},
  author={Canaday, Daniel and Griffith, Aaron and Gauthier, Daniel J},
  journal={Chaos: An Interdisciplinary Journal of Nonlinear Science},
  volume={28},
  number={12},
  year={2018},
  publisher={AIP Publishing}
}

@article{wang2024ultrafast,
  title={Ultrafast silicon photonic reservoir computing engine delivering over 200 TOPS},
  author={Wang, Dongliang and Nie, Yikun and Hu, Gaolei and Tsang, Hon Ki and Huang, Chaoran},
  journal={Nature Communications},
  volume={15},
  number={1},
  pages={p.10841},
  year={2024},
  publisher={Nature Publishing Group UK London}
}

@article{scott1990gain,
  title={Gain and noise characteristics of a Brillouin amplifier and their dependence on the spatial structure of the pump beam},
  author={Scott, AM and Watkins, David E and Tapster, P},
  journal={JOSA B},
  volume={7},
  number={6},
  pages={929--935},
  year={1990},
  publisher={Optica Publishing Group}
}

@article{nathe2023reservoir,
  title={Reservoir computing with noise},
  author={Nathe, Chad and Pappu, Chandra and Mecholsky, Nicholas A and Hart, Joe and Carroll, Thomas and Sorrentino, Francesco},
  journal={Chaos: An Interdisciplinary Journal of Nonlinear Science},
  volume={33},
  number={4},
  year={2023},
  publisher={AIP Publishing}
}

@book{vayshenker2000optical,
  title={Optical fiber power meter calibrations at NIST},
  author={Vayshenker, Igor and Li, Xiaoyu and Livigni, David J and Scott, Thomas R and Cromer, Christopher L},
  journal={Special Publication (NIST SP) - 250-54},
  issue={3},
  year={2000},
  publisher={NIST}
}

@article{dudas2023quantum,
  title={Quantum reservoir computing implementation on coherently coupled quantum oscillators},
  author={Dudas, Julien and Carles, Baptiste and Plouet, Erwan and Mizrahi, Frank Alice and Grollier, Julie and Markovi{\'c}, Danijela},
  journal={npj Quantum Information},
  volume={9},
  number={1},
  pages={p.64},
  year={2023},
  publisher={Nature Publishing Group UK London}
}

@article{ehlers2025stochastic,
  title={Stochastic reservoir computers},
  author={Ehlers, Peter J and Nurdin, Hendra I and Soh, Daniel},
  journal={Nature Communications},
  volume={16},
  number={1},
  pages={1--11},
  year={2025},
  publisher={Nature Publishing Group}
}

@article{zhu2024minimalistic,
  title={Minimalistic and Scalable Quantum Reservoir Computing Enhanced with Feedback},
  author={Zhu, Chuanzhou and Ehlers, Peter J and Nurdin, Hendra I and Soh, Daniel},
  journal={arXiv preprint arXiv:2412.17817},
  year={2024}
}

@article{ehlers2025improving,
  title={Improving the performance of echo state networks through state feedback},
  author={Ehlers, Peter J and Nurdin, Hendra I and Soh, Daniel},
  journal={Neural Networks},
  volume={184},
  pages={107101},
  year={2025},
  publisher={Elsevier}
}

@article{zhu2024practical,
  title={Practical and scalable quantum reservoir computing},
  author={Zhu, Chuanzhou and Ehlers, Peter J and Nurdin, Hendra I and Soh, Daniel},
  journal={arXiv preprint arXiv:2405.04799},
  year={2024}
}

@article{lu2018ultra,
  title={Ultra-low-noise balanced detectors for optical time-domain measurements},
  author={Lu, Qiming and Shen, Qi and Cao, Yuan and Liao, Shengkai and Peng, Chengzhi},
  journal={IEEE Transactions on Nuclear Science},
  volume={66},
  number={7},
  pages={1048--1055},
  year={2018},
  publisher={IEEE}
}

@article{gasmi2016fast,
  title={Fast-and low-noise optical receiver for near-infrared light detection and ranging},
  author={Gasmi, Khaled},
  journal={Optical Engineering},
  volume={55},
  number={9},
  pages={097101--097101},
  year={2016},
  publisher={Society of Photo-Optical Instrumentation Engineers}
}

@article{yu2025chip,
  title={On-Chip Brillouin Amplifier in Suspended Lithium Niobate Nanowaveguides},
  author={Yu, Simin and Zhou, Ruixin and Yang, Guangcanlan and Zhang, Qiang and Zhu, Huizong and Yang, Yuanhao and Xu, Xin-Biao and Chen, Baile and Zou, Chang-Ling and Lu, Juanjuan},
  journal={Laser \& Photonics Reviews},
  pages={2500027},
  year={2025},
  publisher={Wiley Online Library}
}

@article{shaashoua2024brillouin,
  title={Brillouin gain microscopy},
  author={Shaashoua, Roni and Kasuker, Lir and Kishner, Mor and Levy, Tal and Rotblat, Barak and Ben-Zvi, Anat and Bilenca, Alberto},
  journal={Nature Photonics},
  volume={18},
  number={8},
  pages={836--841},
  year={2024},
  publisher={Nature Publishing Group UK London}
}

\appendix

\section{SBS Transcendental Equation Derivation} \label{sec:A}

This section provides the derivation to the transcendental Equation \eqref{eqn:pouttrans} which was numerically solved to simulate the behavior of the SBS activation function. Starting from Equations \eqref{eqn:sbsp1} and \eqref{eqn:sbsp2}

\begin{equation}
    \frac{dP_\mathrm{p}}{dz} = \frac{dP_\mathrm{s}}{dz} \quad \implies \quad P_\mathrm{p}(z) = P_\mathrm{s}(z) + C.
\end{equation}

We can write the integration constant as
\begin{equation}
    C = P_\mathrm{p}(L) - P_\mathrm{s}(L) = P_\mathrm{p} - P_{\mathrm{out}},
\end{equation}
where we define $P_{\mathrm{s}}(L) = P_{\mathrm{out}}$. We also define $P_\mathrm{s}(0) = P_{\mathrm{in}}$. Here, $L$ is the length of SBS interaction. It follows from Equation \eqref{eqn:sbsp2} that
\begin{equation}
    \frac{dP_{\mathrm{s}}}{P_{\mathrm{s}}(P_{\mathrm{s}} + P_{\mathrm{p}} - P_{\mathrm{out}})} = -gdz.
\end{equation}
This can be formally integrated as
\begin{equation}
    \int_{P_{\mathrm{s}}(0)}^{P_{\mathrm{s}}(z)}\frac{dP_{\mathrm{s}}}{P_{\mathrm{s}}(P_{\mathrm{s}} + P_{\mathrm{p}} - P_{\mathrm{out}})} = -\int_0^zgdz'.
\end{equation}
\begin{equation}
    \implies ln\left(\frac{P_{\mathrm{s}}(z)(P_{\mathrm{in}} + P_{\mathrm{p}} - P_{\mathrm{out}})}{P_{\mathrm{in}}(P_{\mathrm{s}}(z) + P_{\mathrm{p}} - P_{\mathrm{out}})}\right) = -gCz.
\end{equation}
This can be rewritten as
\[
    P_{\mathrm{s}}(z)(P_{\mathrm{in}} + P_{\mathrm{p}} - P_{\mathrm{out}}) - P_{\mathrm{in}}P_{\mathrm{p}}e^{-(P_{\mathrm{p}} - P_{\mathrm{out}})gz} 
\]
\begin{equation}
    = (P_{\mathrm{s}}(z)P_{\mathrm{in}} - P_{\mathrm{out}}P_{\mathrm{in}})e^{-(P_{\mathrm{p}} - P_{\mathrm{out}})gz}.
\end{equation}
If we substitute $z = L$ and use $P_\mathrm{s}(L) = P_{\mathrm{out}}$, we get a transcendental equation in terms of $P_{\mathrm{out}}$

\begin{equation}
    P_{\mathrm{out}}(P_{\mathrm{in}} + P_{\mathrm{p}} - P_{\mathrm{out}}) - P_{\mathrm{in}}P_{\mathrm{p}}e^{-(P_{\mathrm{p}} - P_{\mathrm{out}})gL} = 0.
    \label{eqn:pouttrans2}
\end{equation}
This is the required final result.

\end{document}